\documentclass[11pt,a4paper]{emulateapj}
\usepackage{graphicx}
\usepackage{textcomp}
\usepackage{amsmath}
\usepackage{amssymb}
\usepackage{gensymb}
\usepackage{rotating}
\usepackage{epsfig}
\usepackage{amsmath}
\usepackage{longtable}
\usepackage{color}
 \def\apjl{ApJL}

 \def\aap{A\&A}
 
 \def\apjs{ApJS}

 \def\gs{\mathrel{\raise0.35ex\hbox{$\scriptstyle >$}\kern-0.6em\lower0.40ex\hbox{{$\scriptstyle \sim$}}}}
 \def\ls{\mathrel{\raise0.35ex\hbox{$\scriptstyle <$}\kern-0.6em\lower0.40ex\hbox{{$\scriptstyle \sim$}}}}

 \def\Msol{\mathrel{\rm M_{\odot}}}
 \def\Lsol{\mathrel{\rm L_{\odot}}}
 \def\Msolyr{\mathrel{\rm M_{\odot}\,yr^{-1}}}
 \def\Wm2{\,\hbox{W}\,\hbox{m}^{-2}}
 \def\gsim{\mathrel{\raise0.35ex\hbox{$\scriptstyle >$}\kern-0.6em\lower0.40ex\hbox{{$\scriptstyle \sim$}}}}
 \def\lsim{\mathrel{\raise0.35ex\hbox{$\scriptstyle <$}\kern-0.6em\lower0.40ex\hbox{{$\scriptstyle \sim$}}}}
 
 \def\pc{\%}

\lefthead{Simpson et al.}  \righthead{S2CLS: A multi-wavelength study
  of ALMA--identified SMGs in UDS}

\begin{document}

\title{The SCUBA-2 Cosmology Legacy Survey: Multi-wavelength Properties
  of ALMA--identified Submillimeter Galaxies in UKIDSS--UDS}

\author{
J.\,M.\ Simpson,\altaffilmark{1,2}
Ian Smail,\altaffilmark{2,3}
A.\,M.\ Swinbank,\altaffilmark{2,3}
R.\,J.\ Ivison,\altaffilmark{1,4}
J.\,S.\ Dunlop,\altaffilmark{1}
J.\,E.\ Geach,\altaffilmark{5}
O.\ Almaini,\altaffilmark{6}
V.\ Arumugam,\altaffilmark{1,4}
M.\,N.\ Bremer,\altaffilmark{7}
Chian-Chou\ Chen,\altaffilmark{2,4}
C.\ Conselice,\altaffilmark{6}
K.\,E.\,K.\ Coppin,\altaffilmark{5}
D.\ Farrah,\altaffilmark{8}
E.\ Ibar,\altaffilmark{9}
W.\,G.\ Hartley,\altaffilmark{10}
C.\,J.\ Ma,\altaffilmark{2}
M.\,J.\ Micha{\l}owski,\altaffilmark{1}
D.\ Scott,\altaffilmark{11}
M.\ Spaans,\altaffilmark{12}
A.\,P.\ Thomson,\altaffilmark{2}
P.\,P.\ van der Werf \altaffilmark{13}
}

\setcounter{footnote}{0}
\altaffiltext{1}{Institute for Astronomy, University of Edinburgh, Royal Observatory, Blackford Hill, Edinburgh EH9 3HJ, UK; email: jms@roe.ac.uk} 
\altaffiltext{2}{Centre for Extragalactic Astronomy, Department of Physics, Durham University, South Road, Durham DH1 3LE, UK} 
\altaffiltext{3}{Institute for Computational Cosmology, Durham University, South Road, Durham DH1 3LE, UK.}
\altaffiltext{4}{European Southern Observatory, Karl Schwarzschild  Strasse 2, Garching, Germany}
\altaffiltext{5}{Centre for Astrophysics Research, Science and Technology Research Institute, University of Hertfordshire, Hatfield AL10 9AB, UK} 
\altaffiltext{6}{School of Physics and Astronomy, University of Nottingham, Nottingham NG7 2RD, UK}
\altaffiltext{7}{School of Physics, HH Wills Physics Laboratory, Tyndall Avenue, Bristol BS8 1TL, UK}
\altaffiltext{8}{Department of Physics, Virginia Tech, Blacksburg, VA 24061, USA }
\altaffiltext{9}{Instituto de F\'isica y Astronom\'ia, Universidad de Valpara\'iso, Avda. Gran Breta\~na 1111, Valpara\'iso, Chile}
\altaffiltext{10}{Astrophysics Group, Department of Physics and Astronomy, University College London, 132 Hampstead Road, London NW1 2PS, UK}
\altaffiltext{11}{Department of Physics \& Astronomy, University of British Columbia,6224 Agricultural Road, Vancouver, BC, V6T 1Z1, Canada}
\altaffiltext{12}{Kapteyn Astronomical Institute, University of Groningen, The Netherlands}
\altaffiltext{13}{Leiden Observatory, Leiden University, P.O. Box 9513, NL-2300 RA Leiden, Netherlands}

\begin{abstract} 
We present a multi-wavelength analysis of 52 sub-millimeter galaxies (SMGs), identified using ALMA 870\,$\mu$m continuum imaging in a pilot program to precisely locate bright SCUBA2--selected sub-mm sources in the UKIDSS Ultra Deep Survey (UDS) field. Using the available deep (especially near--infrared), panoramic imaging of the UDS field at optical--to--radio wavelengths we characterize key properties of the SMG population. The median photometric redshift of the bright ALMA\,/\,SCUBA--2 UDS (AS2UDS) SMGs that are detected in a sufficient number of wavebands to derive a robust photometric redshift is $z$\,=\,2.65\,$\pm$\,0.13. However, similar to previous studies, 27\,$\pc$ of the SMGs are too faint at optical-to-near--infrared wavelengths to derive a reliable photometric redshift. Assuming that these SMGs lie at z\,$\gsim$\,3 raises the median redshift of the full sample to $z$\,=\,2.9\,$\pm$\,0.2. A subset of 23, unlensed, bright AS2UDS SMGs have sizes measured from resolved imaging of their rest-frame far--infrared emission. We show that the extent and luminosity of the far--infrared emission are consistent with the dust emission arising from regions that are, on average, optically thick at a wavelength of $\lambda_0$\,$\ge$\,75\,$\mu$m (1--$\sigma$ dispersion of 55--90\,$\mu$m). Using the dust masses derived from our optically-thick spectral energy distribution models we determine that these galaxies have a median hydrogen column density of $N_{H}$\,=\,9.8$_{-0.7}^{+1.4}$$\times$10$^{23}$\,cm$^{-2}$, or a corresponding median $V$--band obscuration of $A_\mathrm{v}$\,=\,540$^{+80}_{-40}$\,mag, averaged along the line of sight to the source of their restframe $\sim$\,200\,$\mu$m emission. We discuss the implications of this extreme attenuation by dust for the multiwavelength study of dusty starbursts and reddening--sensitive tracers of star formation. 
\end{abstract}

\keywords{galaxies: starburst, galaxies: high redshift}

\section{Introduction}
\label{sec:intro}
In the local Universe, Ultra Luminous Infrared Galaxies (ULIRGs), with far--infrared luminosities of $\ge$\,10$^{12}$\,$\Lsol$, represent the most intense sites of ongoing star formation (e.g.\ \citealt{Soifer84}). Despite ULIRGs having immense star formation rates of $\gsim$\,100\,$\Msolyr$ the bolometric luminosity of these galaxies is dominated by emission from inter--stellar dust grains, which obscure the ongoing starburst at ultra--violet and optical wavelengths and re-emit in the far--infrared. The dust emission takes the form of a modified blackbody function, which at the typical temperatures of these sources ($T_d$\,$\sim$\,40\,K; e.g.\ \citealt{Symeonidis13}) peaks at $50$--$100$\,$\mu$m restframe and declines strongly at longer wavelengths. The strong increase in flux density with decrease wavelength results in the well--known negative $k$--correction at sub-millimeter (sub-mm) wavelengths; for sources at higher redshift the observed sub-mm waveband samples the spectral energy distribution (SED) closer to its peak. Indeed, the negative $k$--correction is so strong at sub-mm wavelengths ($\sim$\,850\,$\mu$m) that a galaxy with a constant luminosity and temperature has an almost constant brightness with redshift, as the increase in the flux density of the source directly counters the effect of cosmological dimming out to $z$\,$\sim$\,7 (see \citealt{Blain02}). Observations at sub--mm wavelengths thus provide a unique tracer of obscured star formation across a large fraction of the age of the Universe.

The first deep, extragalactic surveys at sub-mm wavelengths, undertaken with bolometer cameras on single--dish facilities, unveiled a population of bright sources at flux densities of $S_{850}$\,$\gsim$\,5--15\,mJy (e.g.~\citealt{Smail97,Barger98,Hughes98,Eales99,Greve04,Coppin06,Scott08}). While these surveys detected only a modest number of sub--mm sources the surface density of these detections was used to infer that the number of far--infrared--bright galaxies must undergo a rapid evolution with redshift (e.g.~\citealt{Smail97}). However, the low angular resolution of single--dish facilities (typically $\sim$\,15$''$) means that identifying the sub-mm galaxies (SMGs; $S_{850}$\,$>$\,1\,mJy) that are responsible for each sub-mm source is not possible without significant assumptions about the properties of these sources at other wavelengths. Typically, the correlation between radio and far--infrared emission (e.g.~\citealt{Ivison98,Ivison00}) was exploited to provide statistical identification of the counterparts to sub-mm sources, since facilities such as the Very Large Array (VLA) can provide the arcsecond resolution imaging that is required to identify individual galaxies \citep{Ivison02,Ivison04,Ivison07,Bertoldi07,Biggs11,Lindner11}. 

Identifying single--dish--detected sub-mm sources through observations with the VLA at 1.4\,GHz paved the way for our understanding of the SMG population, with the initial analysis of these radio--identified SMGs confirming the high--redshift nature (median $z$\,$\sim$\,2.5; \citealt{Chapman05}) and ULIRG--like far--infrared luminosities ($\ge$\,10$^{12}$\,$\Lsol$; e.g. \citealt{Magnelli12}) of the SMG population. Further analysis has shown that SMGs are relatively massive, gas-rich galaxies (gas masses of $\sim$\,5\,$\times$\,10$^{10}$\,M$_\odot$, e.g.\ \citealt{Bothwell13}) with space densities of $\sim$\,10$^{-5}$\,Mpc$^{-3}$ \citep{Hainline11}, while rest--frame optical imaging from {\it HST} has demonstrated that the visible stellar component in SMGs has a disturbed or irregular morphology (e.g. \citealt{Conselice03b,Chapman03b,Swinbank10,Targett13,Wiklind14,Chen14}). Thus, SMGs appear to have similar properties to local ULIRGs, despite being $\sim$\,$10^{3}$ times more numerous than these proposed analogs, at a fixed far--infrared luminosity (e.g. \citealt{Chapman05,Lindner11,Magnelli12,Yun12,Swinbank13}). 

Although identifying sub-mm sources at radio wavelengths has proven to be a powerful tool to understand the SMGs population the technique is susceptible to issues with mis--identification and incompleteness, problems that are inherent in any analysis involving statistical associations. Thus, recent interferometric observations undertaken at sub-mm\,/\,mm wavelengths with facilities such as the Sub-mm Array (SMA) and Plateau de Bure Interferometer (PdBI), or, more recently, with the Atacama Large sub--\,/\,Millimeter Interferometer (ALMA) have improved our understanding of the SMG population (e.g.\ \citealt{Younger09,Wang11,Smolcic12,Hodge13,Karim13,Walter16,Dunlop17}). Crucially, these facilities can provide imaging at sub-mm wavelengths with an angular resolution of only a few arcseconds, or better, thus providing the sub--arcsecond positional accuracy that is required to identify the multi-wavelength counterparts to single--dish identified sub-mm sources and circumventing the requirement for statistical associations at other wavelengths. Furthermore, observations with ALMA can achieve sufficiently high resolution to resolve SMGs at sub-mm wavelengths, enabling the internal processes that govern the obscured starburst to be studied and allowing a direct comparison with local ULIRGs. 

In the first, large interferometric study of sub-mm sources undertaken, \citet{Hodge13} used ALMA to obtained sensitive, high resolution images of 126 sub--mm sources that were identified in the LABOCA survey of the Extended {\it{Chandra}} Deep Field--South (LESS), identifying 99 SMGs within the 17.3$''$--diameter primary beam of 88 of the highest quality ALMA observations. These observations confirmed previous suggestions that a significant fraction of bright single--dish--identified sub-mm sources are comprised of a blend of multiple individual SMGs (e.g.\ \citealt{Wang11}) and led to the suggestion that the intrinsic 870\,$\mu$m number counts may have a strong decline at $>$\,9\,mJy, potentially indicating a maximal luminosity to high redshift starbursts (\citealt{Karim13}; but see also \citealt{Barger12,Chen13}). However, while ALMA--LESS (ALESS) represents a complete sample of sub-mm sources ($S_{870\mu\mathrm{m}}$\,$>$\,4.4\,mJy) the steep shape of the single--dish 870\,$\mu$m number counts means that only 10 sources brighter than $S_{870\mu\mathrm{m}}$\,9\,mJy where observed as part of the survey.

To investigate the properties of the brightest, unlensed SMGs we undertook a pilot ALMA survey of 30 bright sub--mm sources (\citealt{Simpson15b}) that were identified as part of the SCUBA--2 Cosmology Legacy Survey (\citealt{Geach17}). These 30 sources are located in the Ultra Deep Survey (UDS) field, the deepest component of the panoramic UKIRT Infrared Deep Sky Survey (UKIDSS; \citealt{Lawrence07}), and thus have deep multi-wavelength imaging across optical--to--near--infrared wavelengths. In previous work, we presented the source catalog, number counts and far--infrared morphologies of the 52 SMGs that were detected in these 30 ALMA maps (see \citealt{Simpson15,Simpson15b}). We demonstrated that 61$^{+19}_{-15}$\,$\pc$ of single--dish--identified sub-mm are comprised of two or more SMGs ($>$\,1\,mJy) and that the number density of these secondary sources is inconsistent with them being chance line--of--sight projections \citep{Simpson15b}. Furthermore, we used our high--resolution ALMA imaging to show that the far--infrared region in SMGs has a median angular size (deconvolved FWHM of the major axis) of 0.30\,$\pm$\,0.04$''$ \citep{Simpson15}. 

Here, we use the available photometric imaging of the UDS field to study the properties of these 52 ALMA--identified SMGs at optical to radio wavelengths, including an analysis of the dust properties of the 23 SMGs that were resolved in our 870\,$\mu$m ALMA imaging. In particular, the sample of SMGs studied here doubles the number of bright 850\,$\mu$m sources that have been interferometrically--identified using ALMA and we use the improved statistics that this provides to search from trends in the SMG population with flux density. The paper is structured as follows. In \S\,2 we discuss our sample selection. In \S\,3 we describe the multi-wavelength coverage of our ALMA--identified SMGs and our SED fitting procedures, before discussing the multi-wavelength properties of these SMGs in \S\,4. In \S\,5 we present the redshift distribution and far-infrared properties of the AS2UDS SMGs. Furthermore, we discuss the dust properties of the 23 SMGs with measured sizes at observed 870\,$\mu$m and present the implications for the optical depth and attenuation of stellar light in these sources. We discuss these in \S\,5 and give our main conclusions in \S\,6. We adopt a cosmology with $H_{\rm 0}$\,=\,67.8\,km\,s$^{-1}$\,Mpc$^{-1}$, $\Omega_{\Lambda}$\,=\,0.69, and $\Omega_{\rm m}$\,=\,0.31 ~\citep{Planck14}. Throughout this work error estimates are from a bootstrap analysis and all magnitudes are in the AB system \citep{Oke74}, unless otherwise stated.  

\section{Sample Selection}
\label{sec:sample}
In this work we study the multi-wavelength properties of a sample of
52 SMGs that were identified using targeted ALMA Band 7 continuum imaging
of 30 bright single-dish--detected sub-mm  sources. Here we give a brief overview of the sample
selection from the initial single-dish imaging and the ALMA data
reduction. For a detailed description of the data reduction and
analysis we refer the reader to \citet{Simpson15b}. The initial 
sample of 30 sub-mm sources was detected in wide-field SCUBA--2
850\,$\mu$m imaging of the UDS field, taken as part of the SCUBA--2
Cosmology Legacy Survey (S2CLS; \citealt{Geach17}). Our sample was constructed for the ALMA Cycle-1 
deadline in early 2013 from a preliminary version of this S2CLS
map, which reached a 1--$\sigma$ depth of 2.0\,mJy. From this early
map we selected 30 apparently bright sources detected at
$>$\,4\,$\sigma$ for ALMA follow-up observations. Crucially, the ALMA
follow-up observations targeted each sub-mm source at the same
wavelength and provide imaging across a primary beam that encompasses the SCUBA--2 beam (FWHM\,=\,14.8$''$), but with a synthesized beam that is a factor of
$\gsim$\,400 smaller.    

All 30 SCUBA-2--detected sub-mm sources in our sample were observed
with ALMA on 2013 November 1 with 26 12\,m antennae. The array
configuration yielded a synthesized beam (using Briggs weighting with robust\,=\,0.5) of
0.35$''$\,$\times$\,0.25$''$ and the data were calibrated and imaged
using the {\sc Common Astronomy Software Application} ({\sc{casa}};
version  4.2.1). We note that two versions of the final, cleaned maps were produced: a
``high resolution'' set of maps with a median 1--$\sigma$ depth of
0.21\,mJy\,beam$^{-1}$ and a median
synthesized beam of 0.35$''$\,$\times$\,0.25$''$; and a set of $uv$--tapered ``detection'' images with a
median 1--$\sigma$ depth of 0.26\,mJy\,beam$^{-1}$ and a median
synthesized beam of 0.80$''$\,$\times$\,0.65$''$. \citet{Simpson15b} construct a source
catalog of 52 SMGs ($S_{870}$\,=\,1.3-12.9\,mJy) that are detected at $>$\,4\,$\sigma$ in the 30
ALMA ``detection'' images. A subset of 23\,/\,52 SMGs are detected at a sufficiently high SNR ($>$\,10) in the
``high resolution'' images to allow a study of their morphology at
observed 870\,$\mu$m.

\begin{figure*}
  \centering
  \includegraphics[width=0.9\textwidth]{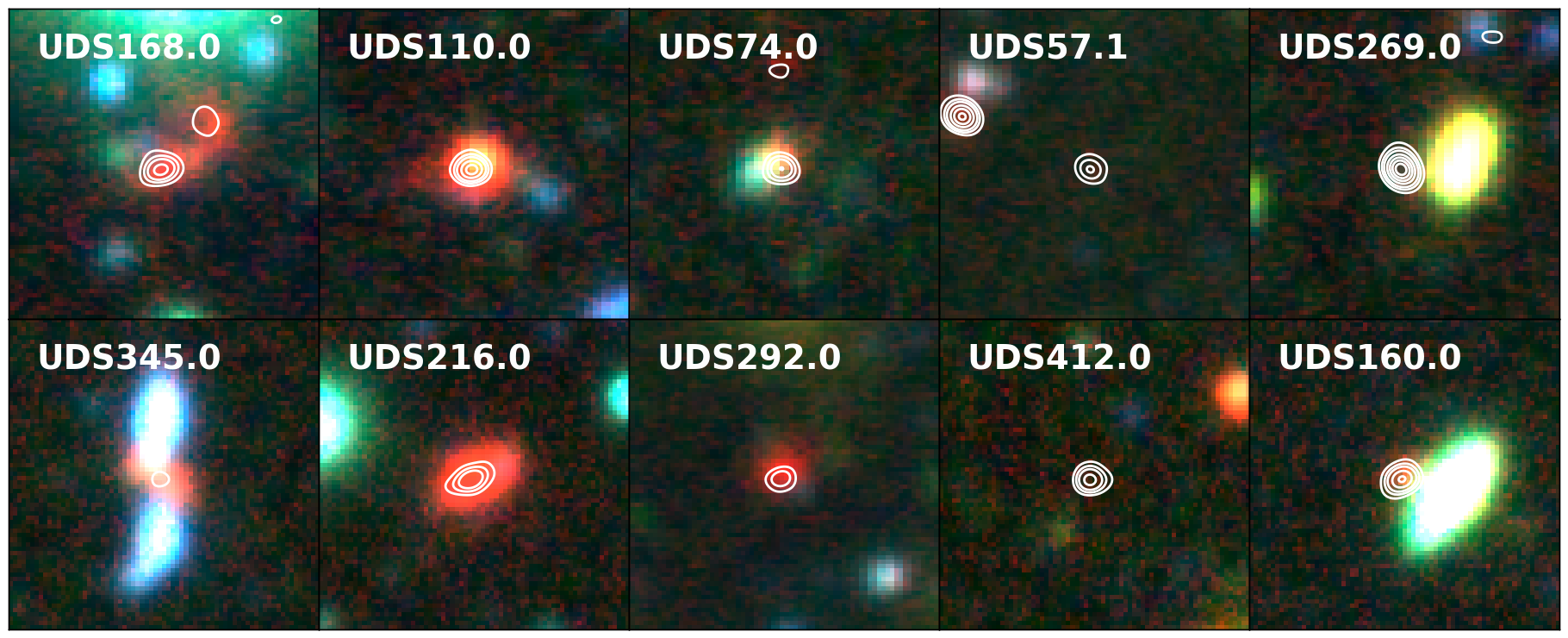} 
  \caption{Example 10$''$\,$\times$10$''$ true color images (constructed from $B$, $I$ and
    $K$) of ten SMGs in our ALMA--identified sample. The sources are
    selected to be representative of the optical-to-near--infrared
    properties of the full sample and from left-to-right the columns show
    sources with disturbed\,/\,irregular morphologies, compact sources
    and optically--blank sources. The final column shows two of the
    four SMGs in our sample that we identify as gravitationally lensed
    sources. The SMGs typically appear red in color, although we
    note that 27\,$\pm$\,7\,$\pc$ of the sample are not detected in the deep
    UKIDSS $K$--band imaging (5--$\sigma$ depth $K$\,=\,24.6).
}
\label{fig:bik}
\end{figure*}

\section{Observations}
\label{sec:obsanalysis}

The focus of this paper is a multi-wavelength analysis of the 52 SMGs detected at SNR\,$>$\,4 in our deep ALMA imaging as part of a pilot study for a large survey of $\sim$\,10$^3$ SMGs with ALMA in the S2CLS UDS map (Stach et al.\ in prep.). Crucially, this pilot AS2UDS sample is comprised of 17 SMGs with 870\,$\mu$m flux densities brighter than 7.5\,mJy, a factor of two increase relative to previous ALMA surveys of 870\,$\mu$m sources (ALESS; \citealt{Hodge13}), and we use this increase in dynamic range to study both bright sources and the overall SMG population ($S_{850}$\,$\gsim$\,1\,mJy). The UKIDSS UDS is a target field for deep, panoramic observations from optical--to--radio wavelengths and we use the existing archival images from these multi-wavelength surveys throughout our analysis. In the following we give a description of each of these surveys and how we use the resulting data products to analyze our sample of SMGs.

\subsection{Optical and Near-Infrared Imaging}
\label{subsec:optnirphot}
The dust enshrouded nature of SMGs means that deep
near-infrared imaging is essential for determining properties such as
their photometric redshifts. The UKIDSS observations of the $\sim$\,0.8\,deg$^{2}$ UDS comprise four
Wide-Field Camera (WFCAM) pointings, in the $J$--,$H$--, and $K$--bands. In this
paper we use the images and catalogs released as part 
of the UKIDSS data release 8 (DR8). The DR8 release contains data
taken between 2005 and 2010, and the final $J$-, $H$-, and $K$-band mosaics
have a median 5--$\sigma$ depth (2$''$ apertures) of $J$\,=\,24.9, $H$\,=\,24.2 and
$K$\,=\,24.6, respectively.  

Deep observations of the UDS have also been taken in the $U$--band with
Megacam at the Canada France-Hawaii Telescope (CFHT) and in the $B$,
$V$, $R$, $i'$, and $z'$--bands with Suprime-cam at the Subaru telescope. The
Megacam\,/\,CFHT $U$--band imaging reaches a $5$\,$\sigma$ (2$''$
diameter aperture) depth of $U$\,=\,26.75 (Foucaud et al.\ in prep.) and the
Suprime-cam imaging has a limiting 3\,$\sigma$ depth of
$B$\,=\,28.4, $V$\,=\,27.8, $R$\,=\,27.7, $i'$\,=\,27.7, $z'$\,=\,26.7 in the $B$, $V$, $R$, $i'$, and 
$z'$--bands, respectively (2$''$ diameter apertures;
\citealt{Furusawa08}). Furthermore, deep {\it Spitzer} data, obtained as
part of the SpUDS program (PI: J.\ Dunlop), provides imaging reaching a
5\,$\sigma$ depth of $m_{3.6}$\,=\,24.2 and $m_{4.5}$\,=\,24.0 at 3.6\,$\mu$m and 4.5\,$\mu$m,
respectively.

The DR8 UKIDSS catalog contains the $U$--to--4.5\,$\mu$m photometry
for $\sim$\,140,000 sources detected in the deep $K$--band image of the 
UDS. For each source, eleven band photometry was determined by running
Sextractor \citep{Bertin96} in ``dual-image'' mode on the images
described above, using the UKIDSS $K$-band image as the detection
image. The flux of each source was measured in a 3$''$--diameter
aperture and to ensure consistent galaxy colors, aperture
corrections that account for source--blending were applied to $U$-band, 3.6\,$\mu$m and 
4.5\,$\mu$m photometry. \citet{Hartley13} use the color--matched photometry to derive photometric redshifts for the sources in the UKIDSS UDS catalog but, to allow a direct comparison to previous studies, we apply a further correction to convert the $3''$ aperture flux measurements to a ``total'' magnitude. We stack 15 isolated stars in the $K$--band image and determine a 3$''$--to--total aperture correction of $-0.2$\,mag, which we apply to the UKIDSS photometry.

\subsubsection{Photometric Redshifts}
\label{subsec:udsredshifts}
Photometric redshifts were determined for the sources in the
UKIDSS UDS DR8 catalog, using the eleven-band
optical-to-near--infrared photometry described in
\S\,\ref{subsec:optnirphot}. The analysis was presented in
\citet{Hartley13} and \citet{Mortlock13}, but we give a summary
here. The template fitting code {\sc eazy} \citep{Brammer08} was used
to fit a library of seven template spectral energy distributions (SEDs) to
the photometry of each $K$--band--selected source in the DR8
release. First, a subset of 2146 spectroscopically-confirmed sources (excluding AGN) were used to calibrate the photometric redshifts, and correct for any zero-point
offsets between the template SEDs and the UDS photometry. The majority
of these spectroscopic redshifts are drawn from the ESO large program
UDSz (Almaini et al.\ in prep) targeting $z$\,$>$\,$1$ galaxies), but a
small number of archival redshifts are included. The redshift of each
spectroscopically-confirmed source was fixed at the spectroscopic
redshift in the SED fitting and the offsets between the template and
observed fluxes were used to iteratively correct the zero-points of
each of the eleven filters. An offset of 0.15\,mag was applied to the
$U$--band photometry and the offsets in all remaining bands
were $\le$\,0.05\,mag. 

The final photometric redshifts for the spectroscopic sample are found to have a median ($z_{\rm phot}$-$z_{\rm spec}$)\,/\,(1+$z_{\rm spec}$)\,=\,0.020, with a 1--$\sigma$ dispersion of 0.031, indicating very good agreement between the redshifts (catastrophic outliers at ($z_{\rm phot}$-$z_{\rm spec}$)\,/\,(1+$z_{\rm spec}$)\,$>$\,0.15 were removed). We note that as AS2UDS SMGs represent a distinct population of highly dust obscured galaxies the accuracy of photometric redshifts for these sources may be lower than estimated for the overall UKIDSS catalog. Indeed, previous studies have shown that for SMGs with comparable photometry the 1--$\sigma$ dispersion in ($z_{\rm phot}$-$z_{\rm spec}$)\,/\,(1+$z_{\rm spec}$) is typically $\lsim$\,0.1 (\citealt{Simpson14}; Danielson et al.\ 2016). Crucially, these studies do not find any bias in the photometric redshifts, and have demonstrated good agreement between the photometric and spectroscopic redshifts of SMGs. Further details regarding analysis and reliability testing of the UKIDSS photometric redshift catalog are given in \citet{Hartley13} and \citet{Mortlock13}.

\begin{figure*}
  \centering 
  \includegraphics[width=0.95\textwidth]{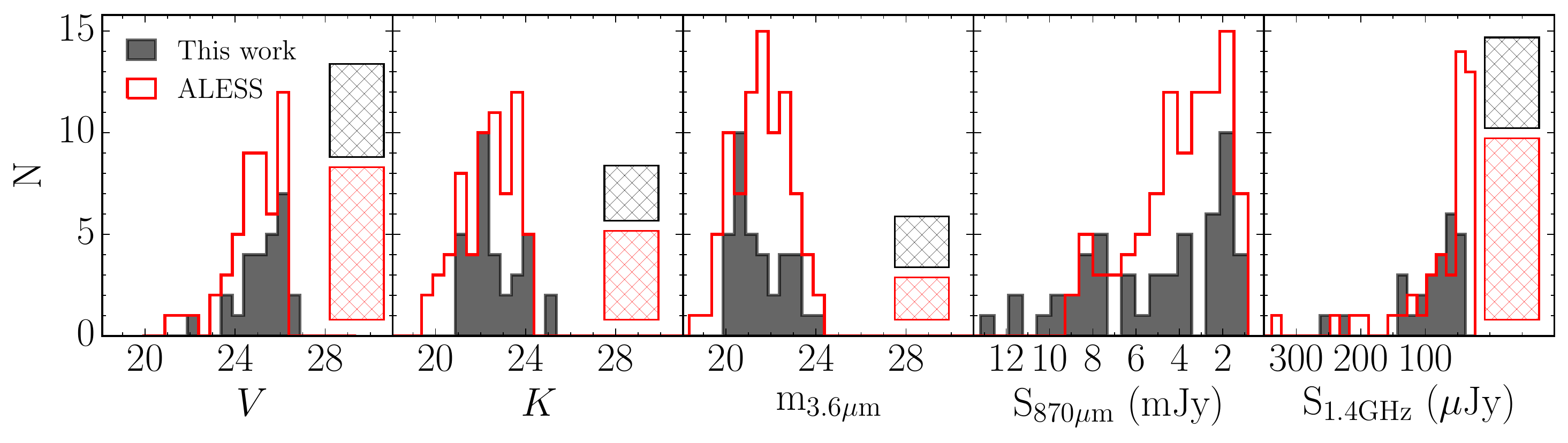}
  \caption{ The apparent magnitude distributions of the AS2UDS sample of SMGs
    in the $V$, $K$ and 3.6\,$\mu$m wavebands, along with their
    flux density distributions at 870\,$\mu$m and 1.4\,GHz. The median
    $V$, $K$ and IRAC 3.6\,$\mu$m apparent magnitudes,
    including the numbers of non--detections (hatched regions), are
    $V$\,=\,$26.4_{-0.3}^{+\infty}$, $K$\,=\,$23.0_{-0.5}^{+0.7}$ and
    $m_{3.6}$\,=\,$21.8_{-0.3}^{+0.6}$. For comparison we 
    show the magnitude distributions of the ALMA-identified SMGs in
    the ALESS sample \citep{Simpson14}. The ALESS SMGs have a median $S_{870}$\,=\,$3.5$\,$\pm$\,0.3\,mJy and so are marginally fainter than the SMGs in our sample, median $S_{870}$\,=\,$4.2^{+0.9}_{-0.6}$\,mJy. The 1.4\,GHz VLA imaging of the UDS reaches
    a 1--$\sigma$ depth of $\sim$\,7\,$\mu$Jy\,beam$^{-1}$, at its deepest, and in total
    27\,/\,52 SMGs from AS2UDS are detected with a median flux density of
    $S_{1.4\mathrm{GHz}}$\,=\,42$^{+11}_{-42}$\,$\mu$Jy (Arumugam et
    al. in prep.). The SMGs in our sample are marginally brighter at 1.4\,GHz
    than the ALESS SMGs (median
    $S_{1.4\mathrm{GHz}}$\,$<$\,19.5\,$\mu$Jy), which we attribute to
    the differences in the 870\,$\mu$m flux density distribution of the two samples.}
\label{fig:hists}
\end{figure*}

\subsection{Far-infrared Imaging}
The UDS field was observed at 250, 350, and 500\,$\mu$m with the
Spectral and Photometric Imaging Receiver (SPIRE) onboard the {\it
  Herschel Space Observatory} as part of the {\it Herschel} Multi-tiered
Extragalactic Survey (HerMES; \citealt{Oliver12}). Observations of the
field were taken in seven ``sub-blocks'', each with an exposure time
of 5.4\,ks, resulting in a total exposure time for the field of
37.8\,ks. As described in \citet{Swinbank13}, we retrieved the Level 2 data products from {\it Herschel}
European Space Agency archive and aligned and co-added the images. To
ensure the co-added SPIRE images are aligned with the astrometric
reference frame of the deep radio imaging of the UDS (see \S\,\ref{subsec:radiodata})
we stacked the maps at the VLA radio positions, centroided the stacked
emission and applied shifts of $<1.5''$ to each SPIRE map.

The SPIRE\,/\,{\it Herschel} imaging has an angular resolution of
$\sim$\,18, 25, and 35$''$ at 250, 350, and 500\,$\mu$m. The
coarse resolution of the imaging means that it is vital that we
consider the effect of source blending when determining the
far-infrared flux densities of the SMGs in our sample. To determine
accurate flux densities for the SMGs we deblend the UDS maps following
the procedure presented in \citet{Swinbank13}, which includes
extensive tests to confirm the reliability and completeness of the
analysis. First, we use the available 24\,$\mu$m\,/\,{\it Spitzer} source
catalogs ($>$\,5\,$\sigma$) to construct a catalog of likely
infrared-bright galaxies that are used as positional priors in the
deblending. The positions for 52 SMGs from ALMA are added to
the prior list and any duplicate sources within 1.5$''$ are removed
from the final prior catalog, ensuring that the ALMA positions are
retained. The SPIRE maps are then deblended by fitting the appropriate beam at the
position of each source in the prior catalog and minimizing the $\chi^{2}$ statistic. To ensure that they do not ``over-deblend''
the longer--wavelength, coarser resolution SPIRE imaging,
\citet{Swinbank13} deblend the maps in order of increasing wavelength
and only include 24\,$\mu$m sources that are detected at
$>2$\,$\sigma$ in the shorter wavebands as positional priors. Upper
limits for non-detections and appropriate error bars are determined
through simulations (see \citealt{Swinbank13}). The detection
fractions of the AS2UDS SMGs are 24\,/\,48, 26\,/\,48, 19\,/\,48, at
250, 350, and 500\,$\mu$m, respectively (25\,/\,48 detected in $\ge$\,2 wavebands), and the deblended SPIRE
fluxes and associated uncertainties are given in
Table\,\ref{table:obs}.  

\subsubsection{Far-infrared SED Fitting}\label{subsubsec:firfitting}
To characterize the temperatures and far--infrared luminosities (8--1000\,$\mu$m) of the SMGs in our
sample we fit the observed far-infrared photometry of each source
with a single--temperature, modified blackbody function

\begin{equation}
\label{eqn:mbb}
  S_{\nu_\mathrm{obs}} \propto (1-e^{-\tau_{\nu_\mathrm{rest}}}) \times B(\nu_\mathrm{rest},T)
\end{equation}

\noindent where $B(\nu,T)$ represents the Planck function, $\tau_{\nu}$\,=\,$(\frac{\nu}{\nu_0})^\beta$ is the frequency dependent
optical depth of the dust, $\nu_0$ is the frequency at which the optical depth is unity, $z$ is the redshift of the source, and $\beta$ is the dust emissivity. In our analysis we adopt $\beta$\,=\,1.8, consistent with previous studies of the far-infrared
emission from SMGs, and in-line with studies of galactic dust emission presented by the {\it Planck} collaboration \citep{Planck11}. 

The optical depth and the dust temperature parameters in the modified blackbody function are correlated; both a decrease in the optical--depth, and an increase in the dust temperature, shifts the peak of the SED blue-wards. To allow a direct comparison to previous work we first make the assumption that the dust emission originates from regions that are optically--thin (i.\,e.\ $\nu_0$\,$\gg$\,$\nu$), simplifying the modified blackbody function to 

\begin{equation}
\label{eqn:opthin}
  S_{\nu_{\mathrm{obs}}} \propto \nu_\mathrm{rest}^{\beta} \times B(\nu_{\mathrm{rest}},T).
\end{equation}

\noindent However, as we discuss in \S\,\ref{subsec:firprops} the emission from SMGs does not originate from regions that are optically--thin, which is consistent with studies of far--infrared--bright sources in the local Universe (e.g. \citealt{Scoville17}). As such, the dust temperature derived using the optically--thin approximation does not represent the true temperature of the dust emission regions and in the following work we refer to it as a characteristic dust temperature. We first compare the characteristic dust temperatures of the AS2UDS SMGs to samples of local sources that have been analyzed in the same manner before estimating the true dust temperatures of these SMGs in \S\,\ref{subsec:firprops}.

We fit the optically-thin modified blackbody function (Equation\,\ref{eqn:opthin}) to the photometry of
each SMG in our sample that has a photometric redshift, using an affine-invariant, Markov Chain Monte
Carlo (MCMC) sampler ({\sc emcee}; \citealt{EMCEE13}). By using an MCMC approach to model the far-infrared emission we can include the full redshift probability distribution function for each SMG, and thus determine robust uncertainties for each model parameter. The MCMC code is run using 50 ``walkers'' for a total 10$^6$ steps, following an initial, and conservative, burn--in phase of 10$^4$ steps. An analysis of the time-series data indicates that for each source
the burn-in phase is complete and the chain is well--mixed. 
As discussed in \S\,\ref{subsec:smgphotometry}, a number of the
SMGs in our sample are not detected in some, or all, of
the SPIRE wavebands. To account for non-detections in the SED fitting
we adopt the modification to the $\chi^{2}$ statistic presented by
\citet{Sawicki12} 

\begin{multline}
\chi^{2}_{\rm mod}\,=\,
\sum_{i}\left(\frac{f_{d,i}-f_{m,i}}{\sigma_i}\right)^2 - \\
2\sum_{j}\textrm{ln} \left\{ \sqrt{\frac{\pi}{2}} \sigma_j  \left[ 1 + {\rm
    erf}\left(\frac{f_{lim,j}-f_{m,j}}{\sqrt{2}\sigma_j}\right)
\right]  \right\}
\end{multline}

\noindent where the summations over $i$ and $j$ represent wavebands in which a
source is detected or non-detected, respectively; $f_d$ is the observed
flux density of a source; $f_m$ is the model flux density; $\sigma$ is the
uncertainty on the measured flux density; and $f_{lim}$ is the upper
limit on the flux density of the source in the relevant waveband. If a
source is detected in all wavebands then the summation over $j$ vanishes
and the statistic reverts to the standard $\chi^{2}$
statistic. However, if a source is not detected in the $j$th waveband then
the modification to $\chi^2$ includes the probability that the source
would be considered a non-detection in the imaging, given the current
value of the model. If an SMG is not detected in any of the SPIRE wavebands then
we can only determine a plausible range for its far-infrared
luminosity, which is determined by the maximum characteristic dust temperature that
is consistent with the SPIRE upper limits and the temperature of the
Cosmic Microwave Background (CMB) at the photometric redshift of the
source. To calculate this range we fix the SED at the measured
870$\mu$m flux density and determine the minimum and maximum parameter
values that produce a model in agreement with all upper limits.

The SED model contains three parameters; the normalization, $N$;
the characteristic dust temperature $T_{d}$; and the redshift of the source,
$z$. The well-known degeneracy between temperature and redshift
means that we cannot constrain both parameters without prior
information \citep{Blain96b}. Thus, we use the full redshift probability distribution for
each source, as determined in the optical-to-near--infrared SED
fitting, as a prior on the redshift.  We note that we place an additional, flat prior on the characteristic dust temperature of each source that ensures that it is higher than the lower limit set by the temperature of the CMB, at the appropriate redshift. Finally, it is well--known a single--temperature modified blackbody function fails to reproduce short wavelength ($\lsim$\,50$\mu$m) dust emission from an infrared-bright galaxy, where emission from increasingly warm dust results in a power law flux distribution \citep{Blain02}. We caution that we do not account for this in our analysis and that a single temperature modified blackbody typically under-estimates the total far-infrared luminosity of a source by 20$\pc$, relative to empirical galaxy template SEDs  (e.g.\ \citealt{Swinbank13}).

\subsection{Radio\,/\,1.4\,GHz Imaging}
\label{subsec:radiodata}
The UDS field was observed by the VLA at 1.4\,GHz as part of the project UDS20
(Arumugam et al.\ in prep.). A total of 14 pointings were used to mosaic
an area of $\sim$\,1.3\,deg$^{2}$ centered on the UDS field. The
observations were taken in A, B, and C--D configuration, yielding a
typical synthesized beam of $\sim$\,1.8$''$ FWHM. The final map reaches a 1--$\sigma$ depth of 7\,$\mu$Jy\,beam$^{-1}$ at its deepest and
$\sim$\,7000 sources are detected across the field at a peak SNR\,$>$\,4. 

We match our ALMA catalog to the 1.4\,GHz catalog and
identify 26 matches within 2$''$ (maximum separation 0.9$''$; expected false matching rate $<$\,0.1$\pc$). However,
two bright SMGs (UDS156.0 and 156.1; $S_{870}$\,=\,8.5 and 9.7\,mJy,
respectively) have a small on-sky separation of 2.3$''$. We inspect
the VLA imaging at the position of these sources and identify a bright
1.4\,GHz source that is centered directly between the position of both
SMGs and extended in the direction of both sources. We estimate the flux density of each SMG by fitting two Gaussian profiles centered at the positions of the
ALMA sources.  

Hence, in total 27\,/\,52 ALMA--identified SMGs are detected in the
deep 1.4\,GHz imaging with flux densities ranging from
40--780\,$\mu$Jy (Table~1). The median flux density of the sample is
weakly constrained at 42$_{-42}^{+11}$\,$\mu$Jy. We note that
the astrometry of the ALMA and VLA images is well-aligned, with
median offsets between the ALMA and VLA source positions of
$-$0.08$^{+0.03}_{-0.02}$$''$ in R.\,A.\ and $-$0.03$^{+0.05}_{-0.03}$$''$ in Decl.

\section{Analysis}
We first study the fundamental observable characteristics of our SMG sample before determining their redshifts, which allow us to determine key physical properties such as the epoch of their activity. An extensive literature search reveals that none of the SMGs in our sample have archival spectroscopic redshifts (including from UDSz; \S\,\ref{subsec:udsredshifts}). However, we can make use of the excellent multi-wavelength imaging that is available in the UDS (see \S\,\ref{subsec:optnirphot}) and the photometric redshift estimates that have been derived from the UKIDSS UDS \citep{Hartley13}. In the following section we present the multi-wavelength properties of our sample of AS2UDS SMGs and compare these to other samples of ALMA--identified SMGs. 

\begin{figure}
  \centering
  \includegraphics[width=0.95\columnwidth]{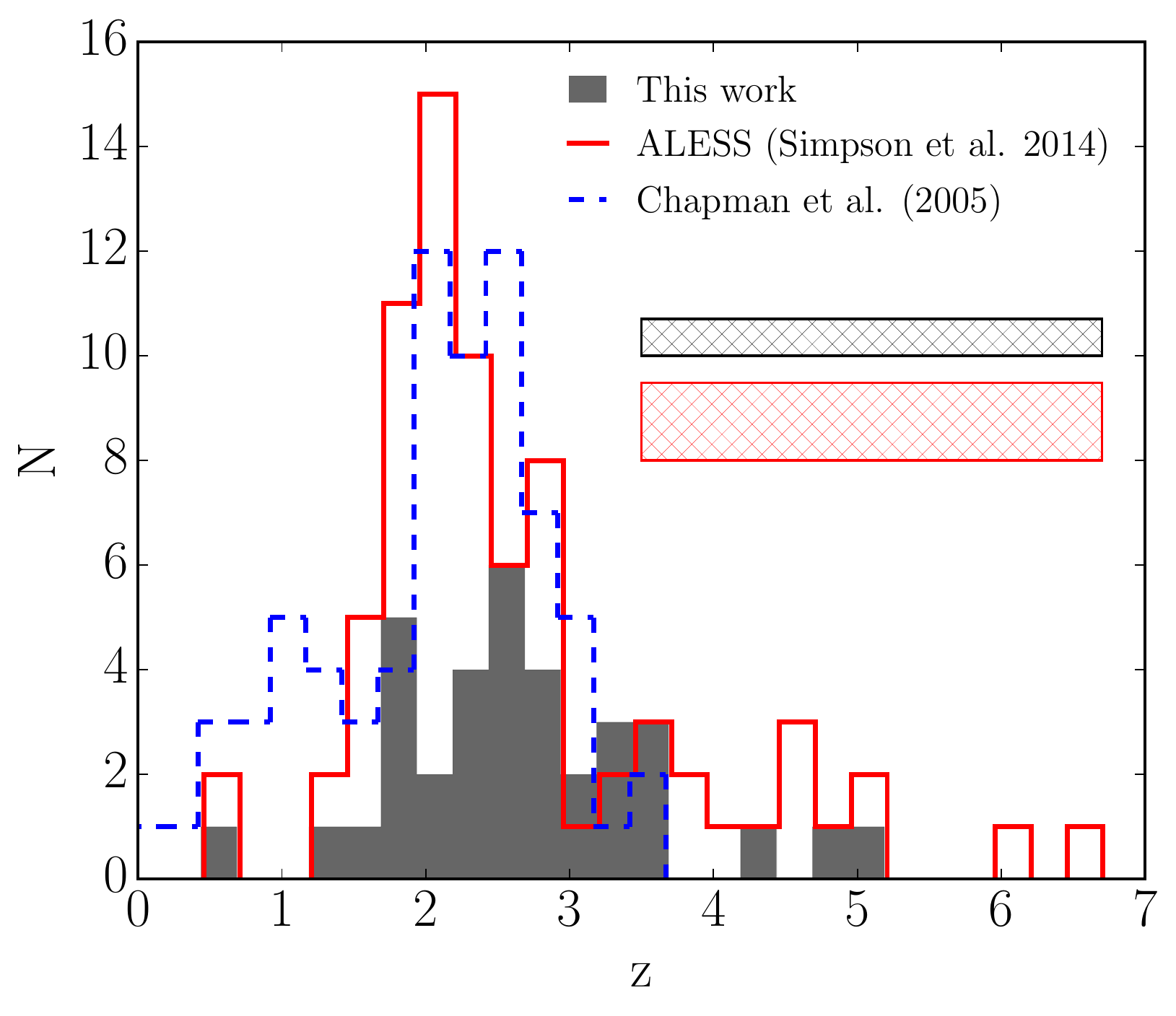}
  \caption{The photometric redshift distribution of the
    ALMA-identified SMGs in our sample. The 35 SMGs in AS2UDS that have
    sufficient photometry to derive a photometric redshift have a
    median redshift of $z$\,=\,2.65\,$\pm$\,0.13. For comparison we
    show the photometric redshift distribution of ALMA--identified SMGs
    in the ECDF--S (ALESS;~\citealt{Simpson14}) and the spectroscopic redshift distribution
    of radio--identified SMGs presented by \citet{Chapman05}. We
    find that the median redshift of the SMGs in our sample is marginally higher than for the ALESS
    SMGs, $z$\,=\,2.31$^{+0.08}_{-0.13}$. However, the median values are
    consistent at the 1.5--$\sigma$ confidence level, and the shape of
    the distributions appear to be in agreement. Similarly, the radio--identified sample
    presented in \citet{Chapman05} lie at a lower median redshift
    of $z$\,=\,2.20\,$\pm$\,0.10, and have notably more sources at
    $z$\,$<$\,1. Hatched regions represent the 13 and 19 SMGs in the AS2UDS and ALESS samples, respectively, that have insufficient photometry to derive a reliable photometric redshift.
}
\label{fig:zdist}
\end{figure}

\subsection{Optical and Near--infrared Photometry}
\label{subsec:smgphotometry}
To determine the optical-to-near--infrared photometry of the SMGs in
our sample we match the ALMA--identified positions to the 
UKIDSS $K$--band catalog. A matching radius of 1$''$ was adopted (85\,\% of matches are found within $\lsim$\,0.5$''$) to
account for both the formal uncertainty on the ALMA positions ($\sigma$\,$\sim$\,0.14$''$ for a 4\,$\sigma$ detection; see \citealt{Ivison07}) and any intrinsic spatial offset resulting from dust obscuration ($\sigma$\,$\sim$\,$0.3''$, with offsets of up to 2$''$ to individual components; \citealt{Chen15}). To ensure that the ALMA and UKIDSS astrometric reference frames are well-aligned, we compare the positions of the 33 matched sources in both catalogs. We identify a small astrometric offset between the reference frames of 0.09$^{+0.05}_{-0.04}$$''$ and $-$0.15$^{+0.04}_{-0.07}$$''$ in R.A. and Decl., which we apply to the UKIDSS UDS astrometry. Note that we then repeated the source matching using the astrometrically aligned catalog, but did not identify any further matches to the sources in our sample at $<$\,1$''$.

To ensure that we have not missed any potential counterparts to the AS2UDS SMGs we extend the search radius for counterparts to $2$\,$''$; consistent with previous high resolution studies of SMGs that have demonstrated significant positional offsets between the observed 870\,$\mu$m and near--infrared emission in a fraction of counterparts as a result of the high dust obscuration, disturbed morphology and often structured dust regions that are typical of the SMG population (e.g. \citealt{Chen15}). Matching the ALMA and UKIDSS catalogs, we identify a potential counterpart to both UDS\,199.1 and UDS\,269.1, at separations of 1.3$''$ and 1.6$''$, respectively. To test the reliability of these proposed counterparts we first construct a catalog of 50,000 random positions within the area of the UKIDSS $K$--band image. We match our fake source list to the UKIDSS source catalog and estimate a false--matching--rate of 8\,\% and 12\,\% at 1.3$''$ and 1.6$''$, respectively. However, previous studies of the redshift distribution of SMGs have indicated that the majority of sources lie at $z$\,$>$\,1.5 and we can use this prior knowledge in our analysis. Thus, we repeat our analysis and estimate that at a separation of 1.3$''$ and 1.6$''$ the false--matching--rate of a source in our catalog of random positions to a $z$\,$>$\,1.5 source in the UKIDSS catalog is 3\,\% and 4\,\%, respectively. Both of the proposed counterparts to both UDS\,199.1 and UDS\,269.1 lie at $z$\,$>$\,1.5 (see Table 2) thus, given the low likelihood that these are spurious matches, we include both in our analysis. 

In Figure~\ref{fig:bik} we show example $BIK$ true--color images for ten SMGs that span the full range of ALMA 870\,$\mu$m flux density for our catalog. The images demonstrate that if an SMG is detected in the optical--to--near-infrared imaging it typically appears red in the $BIK$ color images. The observed $V,$ $K,$ and 3.6\,$\mu$m magnitude distributions of the SMGs in our sample are shown in Figure~\ref{fig:hists}.

It is important to note that the counterparts to the SMGs are
identified by matching to a $K$--band--selected catalog. The depth of
the $K$--band image relative to the IRAC imaging (5\,$\sigma$ depths of $K$\,$=$\,24.6 and $m_{3.6}$\,=\,24.2\,mag) means that we do not expect to have missed a significant number of additional counterparts in the longer wavelength imaging, except for the very reddest sources. Indeed,
we examine the IRAC imaging and only identify counterparts to a further
four SMGs at 3.6\,$\mu$m and\,/\,or 4.5\,$\mu$m (UDS\,57.1, UDS\,199.0, UDS\,286.2, and UDS\,412.0). However, these
sources are not detected at any other wavelengths and, as shown in
\citet{Simpson14}, detections in at least four wave--bands are required
to determine even crude photometric redshifts; a crucial first step towards understanding the physical properties of these sources. We note that three of these SMGs are not detected in the available 1.4\,GHz imaging, and that the far-infrared emission from all four SMGs appears to peak redwards of 350\,$\mu$m, indicating that they likely lie at higher redshift ($z$\,$\gsim$\,3; see \citealt{Swinbank13}).

As our ALMA observations targeted bright sub-mm sources ($S_{850}$\,$\gsim$\,8\,mJy) we must be
aware of the influence of gravitational lensing on our initial
selection (e.g.\ \citealt{Blain96a, Chapman02}). To quantify the effect
of gravitational lensing on our sample we visually inspected the
optical imaging of all 52 SMGs, identifying four sources (UDS\,109.0,
160.0, 269.0 $\&$ 286.0) as being potentially gravitationally
lensed. All four of these SMGs lie close to, but are spatially offset
from, galaxies at $z$\,$<$\,1 (see Figure~\ref{fig:bik}). These SMGs are faint, or undetected, at optical wavelengths, relative to the foreground sources, and although the emission in the IRAC imaging typically appears extended from the bright galaxy in the direction of the SMG it is heavily blended. None of the SMGs show evidence of being multiply imaged indicating that the potential magnification factors are likely to be modest. We highlight these four sources in Table\,\ref{table:obs} and do not include them in our main analysis.   

The median apparent magnitudes of the sample are $V$\,=\,$26.4_{-0.3}^{+\infty}$,
$K$\,=\,$23.0_{-0.5}^{+0.7}$, and $m_{3.6}$\,=\,$21.8_{-0.3}^{+0.6}$. Excluding gravitationally lensed
sources, 27\,$\pm$\,7\,$\pc$ of the sample (13\,/\,48 SMGs) are undetected in the deep
UKIDSS UDS imaging ($K$\,$\le$\,24.6\,mag). As expected for 
dusty, high redshift sources the counterpart detection rate decreases
in bluer wavebands, falling to 54\,$\pm$\,8\,$\pc$ (26\,/\,48) in
the $B$--band. For comparison, in Figure~\ref{fig:hists} we show the magnitude distributions for the 96 ALESS SMGs (\citealt{Simpson14}). The ALESS SMGs \citep{Hodge13} were identified in ALMA 870\,$\mu$m follow--up imaging of single--dish--identified 870\,$\mu$m sub-mm sources and are well--matched to the sample presented here. The parent sample for the AS2UDS SMGs is brighter at 870\,$\mu$m than the ALESS SMGs, and this is reflected in the 870\,$\mu$m flux densities of the sources (median $S_{870}$\,=\,4.2$_{-0.6}^{+0.9}$\,mJy and $S_{870}$\,=\,3.5\,$\pm$\,0.3\,mJy for AS2UDS and ALESS, respectively). The ALESS SMGs have median apparent magnitudes of $V$\,=\,$26.1_{-0.1}^{+0.2}$,
$K$\,=\,$23.0_{-0.4}^{+0.3}$, and $m_{3.6}$\,=\,$21.8_{-0.1}^{+0.2}$, respectively, in
good agreement with the observed magnitude distributions of the AS2UDS SMGs. 

\subsection{Optically--faint SMGs}
We next investigate whether the detectability of counterparts to SMGs in the $K$--band is a function of 870\,$\mu$m flux density. The
$K$--band detected sources in our sample have a median $S_{870}$\,=\,$4.2_{-0.6}^{+1.0}$\,mJy, compared to a median $S_{870}$\,=\,$2.3_{-0.2}^{+1.9}$\,mJy for the non-detections; a small hint, albeit statistically insignificant, that the $K$--band non-detections may be
fainter at 870\,$\mu$m. To investigate this further we combine the AS2UDS and ALESS samples and repeat the
analysis, but to ensure a fair comparison we consider any AS2UDS SMGs fainter than detection limit of the $K$--band imaging of the ALESS SMGs ($K$\,$\le$\,24.4) as non-detected. The median 870\,$\mu$m flux densities for the $K$--band
detections and non-detections in the combined sample are
$S_{870}$\,=\,$4.0$\,$\pm$\,0.3\,mJy and $S_{870}$\,=\,$2.3_{-0.2}^{+0.3}$\,mJy, respectively,
again suggesting that the $K$--band undetected SMGs are fainter at 870\,$\mu$m at the 2.8\,$\sigma$ significance-level. If this
result is confirmed in larger samples then these fainter SMGs represent either the
lower--luminosity (either due to higher dust obscuration or lower stellar
mass) and\,/or high redshift tail to the SMG population. As discussed by \citet{Simpson14} placing these SMGs at low redshift introduces a strong bi-modality into the distribution of rest--frame $H$--band luminosity (a proxy for stellar mass) or dust obscuration in the SMG population. This problem can be avoid by instead assuming that these sources simply represent the high-redshift tail to the SMG population that lie below the detection threshold of the optical--to--near-infrared imaging. Hence, in \S\,\ref{subsec:discphots} we discuss the impact of placing these SMGs at high redshift.

\section{Results $\&$ Discussion}

\begin{figure}
  \centering
  \includegraphics[width=0.95\columnwidth]{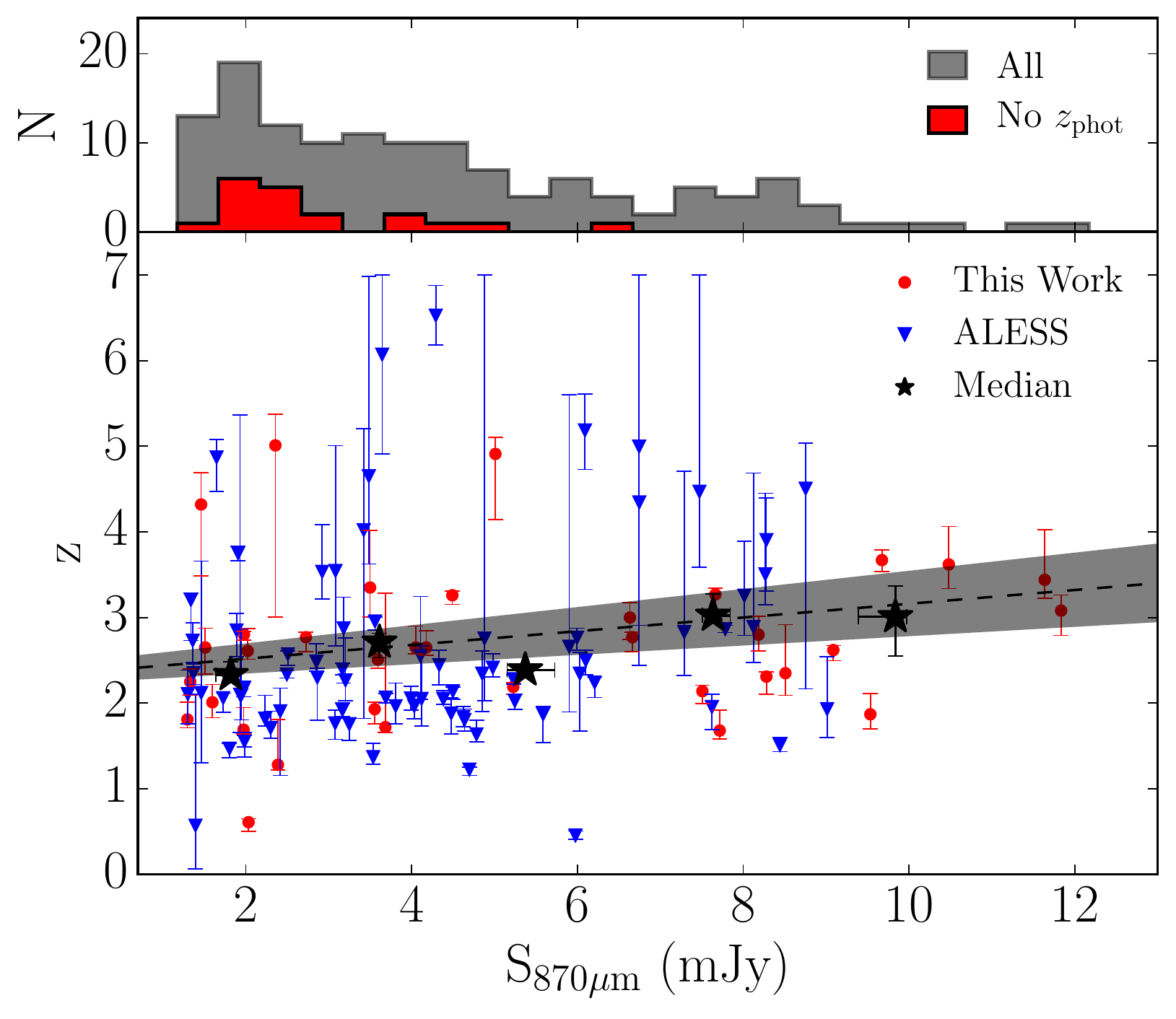}
  \caption{ The photometric redshifts of the 35 SMGs presented in this
    work as a function of their 870\,$\mu$m flux densities. For comparison
    we also show the 77 ALESS SMGs with photometric redshifts detected in ALMA
    imaging of single--dish sources in the ECDF--S
    \citep{Hodge13, Simpson14}. We combine both samples of SMGs and plot the median of the combined sample in 2\,mJy--wide bins. A trend of
    increasing flux density with increasing redshift is observed for
    the SMGs with photometric redshift estimates, and indeed a linear fit to the data shows a slope of 0.080$\pm$0.026 (dashed line and shaded region represent the best-fit and 68$\pc$ confidence region respectively). However, in the 
    upper panel we show the flux density distribution for the SMGs 
    that do not have a photometric redshift estimates and the
    overall sample. If we assume that these optically--faint SMGs lie at $z$\,$>$\,3.0 (a likely hypothesis; \citealt{Simpson14}) then the observed trend in 870$\,\mu$m flux density with redshift weakens, yielding a best-fit slope 0.000$\pm$0.001, and thus is consistent with no evolution with cosmic time. We therefore conclude that there is currently no evidence for a trend of redshift with 870\,$\mu$m flux density for SMGs.
}
\label{fig:s870z}
\end{figure}

\subsection{Photometric Redshift Distribution of SMGs}
\label{subsec:discphots}
The 35 SMGs from our sample of 48 that are detected in the $K$--band imaging
of the UDS have a median redshift of $z_{\mathrm{phot}}$\,=\,2.65\,$
\pm$\,0.13. The shape of the redshift distribution is slightly skewed to high redshift and
extends to $z$\,$\sim$\,5 (Figure~\ref{fig:zdist}) \footnote{A number of the AS2UDS SMGs have large redshift uncertainties, or secondary minima in their redshift probability distribution functions. To investigate whether the overall redshift distribution is sensitive to these we create a single redshift probability distribution for the sample by co-adding the integral--normalized redshift probability function of each SMG. The shape of the combined redshift probability function is well--matched to the shape of the redshift distribution shown in the Figure~\ref{fig:zdist}, and corresponds to a median redshift of $z_{\mathrm{phot}}$\,=\,2.61$^{+0.07}_{-0.13}$, in agreement with the median redshift of the AS2UDS SMGs.}. We first compare the redshift distribution of the AS2UDS SMGs to a sample of radio--identified sub-mm sources with spectroscopic redshifts presented by \citet{Chapman05}. The \citet{Chapman05} sample of SMGs lie at a median redshift
of $z$\,=\,2.20\,$\pm$\,0.10, slightly lower than the redshift of the SMGs
presented here. An offset between the redshift distribution of the
radio--selected and 870\,$\mu$m--selected SMGs is expected due to the
respective positive and negative $k$--corrections in each wave--band. To ensure a fair comparison, we consider the
21\,/\,35 SMGs in our redshift distribution that are detected in the
VLA 1.4\,GHz imaging presented here, which we note has a comparable depth to radio imaging employed by Chapman et al.\ (2005; 7\,$\mu$Jy\,beam$^{-1}$ here versus $\sim$\,9\,$\mu$Jy\,beam$^{-1}$). These radio--detected, ALMA--identified SMGs have a median
redshift of $z_{\mathrm{phot}}$\,=\,2.62$_{-0.31}^{+0.15}$, slightly higher than the sample presented by \citet{Chapman05}, but consistent at the $1$--$\sigma$ confidence level. We note that the median redshift of the radio--identified subset of the AS2UDS SMGs sample is consistent with the $K$--band detected subset, indicating that for the SMGs presented here the radio selection limit is well--matched to the depth of the $K$--band image.

Next, we use the photometric redshifts that we have determined for the AS2UDS SMGs to test whether multiple sources that are detected in the same ALMA map tend to lie at the same photometric redshift. Thus testing if these SMGs are physically--associated or are simply line--of--sight projections. Due to the large associated uncertainties on the photometric redshift of any individual SMG (median $\Delta$\,$z$\,$\sim$\,0.4) we cannot test whether the SMGs located in the same map are physically associated on a source-by-source basis. Instead, we sample the full redshift probability distribution for each SMG and search for statistical over-densities of sources at the same redshift in each ALMA map, relative to the overall population. We find that the AS2UDS SMGs that are detected in the same ALMA map are 17\,$\pm$9\,$\pc$ more likely to lie at $\Delta$\,$z$\,$<$\,0.4, compared to SMGs that are detected in a different ALMA map. While this provides tentative evidence that a fraction of these SMGs are physically--associated we caution that this is a 2\,$\sigma$ results, and that the test can only be performed for the eleven pairs where photometric redshifts are available for both SMGs.

In Figure~\ref{fig:zdist} we compare the redshift distribution of the
AS2UDS SMGs to the photometric redshift distribution of the 77 ALESS SMGs
presented by \citet{Simpson14}. The ALESS SMGs lie at a median
redshift of $z$\,=\,2.3\,$\pm$\,0.1 and we note that the shape of the
distribution is similar to the results presented here; there is a
dearth of SMGs in both samples at $z$\,$\lsim$\,1, and a high redshift
tail extends to $z$\,$>$\,3. A further 19 ALESS
SMGs are detected in an insufficient number of
optical--to--near--infrared wave-bands to determine a photometric
redshift.  The fraction of SMGs in our sample without photometric
redshift estimates is 27$^{+10}_{-7}$\,$\%$ (13\,/\,48), which is consistent with
that for the ALESS sample (20\,$\pm$\,5\,$\%$) at the $<$\,1--$\sigma$ confidence level,
assuming Poisson statistics.   

The median redshift of the SMGs presented in this work is marginally higher
than the ALESS SMGs. The key difference between the samples is that the AS2UDS SMGs are brighter, on average, at 870$\mu$m than the ALESS sample, and have a significantly higher fraction of more luminous sources (29$^{+8}_{-7}$\,$\pc$ at S$_{870}$\,$>$\,7.5\,mJy, compared to
9$^{+4}_{-3}$\,$\pc$ for the ALESS SMGs; see Figure~2). Thus, a possible explanation
for the higher median redshift of the SMGs presented here, relative to
ALESS, is that brighter SMGs are preferentially found at higher
redshift. Indeed, a number of authors have previously suggested that
870--$\mu$m--brighter sources may lie at higher redshift  (e.g.\ \citealt{Ivison02, Ivison07, Koprowski14}). 

To investigate whether there is evidence for such a trend we
combine the AS2UDS and ALESS SMGs, and analyze the combined sample of 144
sources. As shown in Figure~\ref{fig:s870z} we find the SMGs {\it that have photometric redshift estimates} do exhibit a positive trend of
increasing flux density with redshift and a linear fit to the data finds a slope of 0.080\,$\pm$\,0.026. However, we strongly caution
that this trend is mirrored by a decrease in the 
redshift completeness with decreasing 870--$\mu$m
flux; 22\,/\,32 of the SMGs that do not have a photometric
redshift have $S_{870}$\,$<$\,3\,mJy. 

As discussed previously, the optically--faint SMGs that do not have a photometric redshift estimate are likely to lie at higher redshifts than the average AS2UDS SMGs. So, if these SMGs are conservatively placed at $z$\,=\,3--6 then the positive trend between $S_{870}$ and redshift is no longer apparent, and a linear fit to the data returns a slope of $-$0.000\,$\pm$\,0.001. Placing these optically--faint SMGs at $z$\,$>$\,3.0 does however raise the median redshifts of the AS2UDS and ALESS samples to $z$\,=\,2.9\,$\pm$\,0.2 and $z$\,=\,2.5\,$\pm$\,0.2, respectively. As such the median redshift of the AS2UDS SMG is $\Delta$\,$z$\,$\sim$\,0.4 higher than that found for the ALESS SMGs, when non-detections are treated in the same manner. As discussed above, this disparity in the redshift distribution of these samples of SMGs is not due to a difference in the flux density distribution of both samples. Instead, it probably indicates that there is a difference in the underlying distribution of galaxies in the ECDF--S and UDS fields, reinforcing the conclusion that the redshift distribution of SMGs is sensitive to the large-scale-structure of the Universe (\citealt{Williams11}).

\begin{figure*}
  \includegraphics[width=0.97\columnwidth]{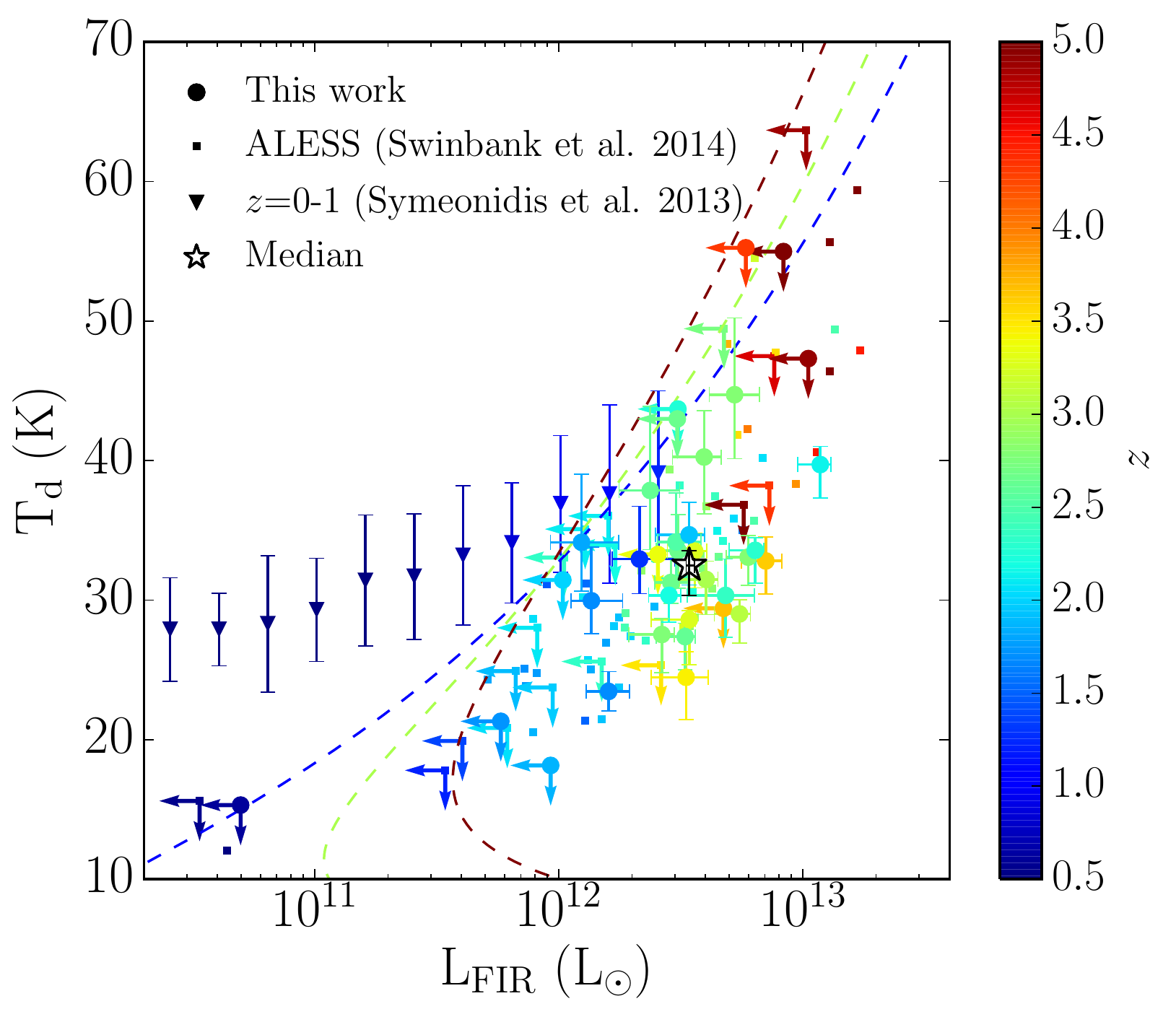}
  \hfill
  \includegraphics[width=0.97\columnwidth]{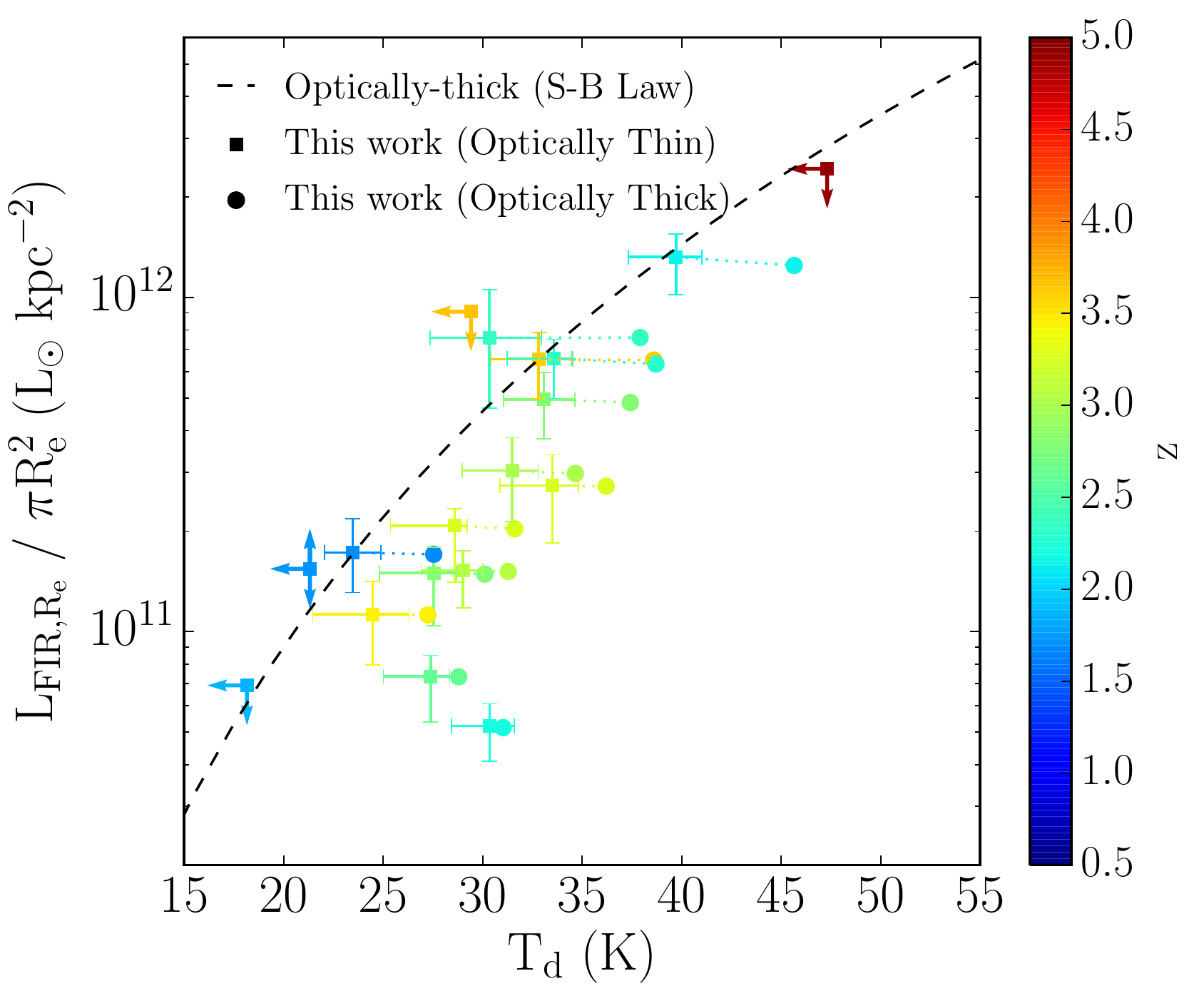}
  \caption{{\it{Left:}} The characteristic dust temperature of the A2SUDS SMGs as a function of far-infrared luminosity, and color-coded by photometric redshift. For comparison we show the ALESS sample \citep{Swinbank13} and a sample of infrared-bright sources at $z$\,$<$\,$1$, selected from observations with {\it Herschel}\,/\,SPIRE \citep{Symeonidis13}. The AS2UDS show a clear trend of increasing characteristic dust temperature with increasing luminosity, consistent with observations of low redshift LIRGs\,/\,ULIRGs, but appear cooler at a fixed luminosity. We caution that due to our selection at 870\,$\mu$m, and the depth of our ALMA observations, the samples do not overlap significantly in far-infrared luminosity; the selection function of our observations at 870\,$\mu$m ($S_{870}$\,$\ge$\,1.4\,mJy), as a function of redshift, is represented by dashed lines for $z$\,=\,1, 3, 5. Nevertheless, the AS2UDS SMGs appear to have characteristic dust temperatures that are $\sim$\,8\,K cooler than sources with comparable far-infrared luminosities at $z$\,$\lsim$\,1. The lower characteristic dust temperatures of the AS2UDS SMGs, at a fixed luminosity, suggest that the dust emission from these sources arises from regions with significantly different physical properties to low redshift far-infrared--bright galaxies, emphasizing that these sources cannot be simply used as analogues to describe high redshift SMGs. {\it{Right:}} The luminosity density of the 18 AS2UDS SMGs that have both photometric redshifts and measured sizes from high resolution 870\,$\mu$m imaging with ALMA. The data show a clear trend of increasing luminosity density with dust temperature, consistent with the Stefan-Boltzmann law for blackbody emission (dashed line) if the dust regions in the sources are assumed to be optically-thin. However, the Stefan-Boltzmann law is only valid if the far-infrared emission originates from dust clouds that are optically--thick at all far-infrared wavelengths. We show that the AS2UDS SMGs are optically-thick at a median $\lambda_0$\,$>$\,75\,$\mu$m, with a 1--$\sigma$ dispersion of 55--90\,$\mu$m, resulting in a systematic increase in the implied intrinsic dust temperatures of each source by $\ge$\,3.1$^{+1.0}_{-0.3}$\,K, on average.}
  \label{fig:lumden}
\end{figure*}

\subsection{Far--infrared Properties}\label{subsec:firprops}
As described in \S\,\ref{subsubsec:firfitting} we estimate the far-infrared luminosities and characteristic dust temperatures of the AS2UDS SMGs by fitting an optically-thin modified blackbody to the observed photometry of each source. In total, 24 AS2UDS SMGs are detected at a sufficient number of optical--to--far--infrared wavelengths that we can estimate both their far-infrared luminosities and characteristic dust temperatures (i.e.\ detected in at least one SPIRE waveband and have a photometric redshift) and these SMGs have a median far--infrared luminosity and characteristic dust temperature of $L_{\mathrm{FIR}}$\,=\,$3.4^{+0.2}_{-0.2}$\,$\times$10$^{12}$\,$\Lsol$ and $T_{\mathrm{d}}$\,=\,$32.9^{+0.4}_{-1.6}$\,K, respectively. In Table\,2 we provided the range of plausible far--infrared luminosities for the 11 AS2UDS SMGs that are not detected in the SPIRE imaging, but have photometric redshift estimates. We compare the far-infrared luminosities and characteristic dust temperatures of the sources presented here to the ALESS sample of SMGs but, to ensure an accurate comparison, we repeat the far-infrared SED fitting for these ALESS SMGs using the photometry presented by ~\citet{Swinbank13} and the SED fitting method presented here. The 59 ALESS SMGs that are detected in at least one SPIRE waveband have a median far--infrared luminosity and characteristic dust temperature of $L_{\mathrm{FIR}}$\,=\,$3.2^{+0.3}_{-0.7}$\,$\times$10$^{12}$\,$\Lsol$ and $T_{\mathrm{d}}$\,=\,$32.1^{+1.3}_{-1.0}$\,K, respectively, consistent with the median properties of the AS2UDS SMGs. The similarity between the median luminosities and the characteristic dust temperatures of the AS2UDS and the ALESS SMGs is unsurprising, given the close agreement between the median 870\,$\mu$m flux density of the samples and the relevant depth of the SPIRE multi-wavelength coverage, but confirms the homogeneity in the properties of bright, 870--$\mu$m--selected sources.

In Figure~\ref{fig:lumden}, we investigate the relationship between the luminosity and characteristic dust temperature of the AS2UDS SMGs and find a clear trend of increasing luminosity with temperature. However, before comparing this result to other samples we must consider the selection function of our 870\,$\mu$m ALMA observations. Thus, in Figure~\ref{fig:lumden} we also show the far-infrared selection function at the depth of our 870\,$\mu$m observations, as a function of both characteristic dust temperature and redshift. We see the well-known negative $k$--correction in the sub-mm waveband that results in near--uniform selection in far-infrared luminosity out to $z$\,$\sim$\,6, for sources at a fixed characteristic dust temperature of $\sim$\,30\,K. However, this uniform selection with redshift does not hold for all dust temperatures. Crucially, at the median redshift of our sample our observations are $\sim$\,3\,$\times$ less sensitive to sources that have a characteristic dust temperature of $T_{d}$\,=\,40\,K, relative to sources with $T_{d}$\,=\,30\,K (see also Blain et al.\ 2002). As such care must be taken when comparing the results presented here to samples selected at a different observed wavelength and\,/\,or redshift.

It has been suggested that SMGs may be the high redshift analogues of ULIRGs that are seen in the local Universe. To investigate whether low-redshift populations can be used as templates for SMGs, we compare the far-infrared properties of the AS2UDS SMGs to the SPIRE--selected sample of $z$\,$\sim$\,0--1 U\,/\,LIRGs presented by \citet{Symeonidis13}. We convert the far-infrared luminosities from \citet{Symeonidis13} to account for the difference between modeling the emission with empirical templates and a modified blackbody, but note that the characteristic dust temperatures are derived in the same manner as the present work. The local sample shows a clear trend of decreasing peak SED wavelength, or equally increasing characteristic dust temperature, with far--infrared luminosity, and the median characteristic dust temperature of the sample rises from $T_{d}$\,=\,29--39\,K over the range $L_{\mathrm{FIR}}$\,=\,0.1--2.5$\times$10$^{12}$\,$\Lsol$. The AS2UDS SMGs exhibit a similar trend between far--infrared luminosity and characteristic dust temperature, relative to the local sample, but appear to be significantly cooler at a fixed luminosity; the AS2UDS SMGs with far--infrared luminosities of $L_{\mathrm{FIR}}$\,=\,3--5$\times$10$^{12}$\,$\Lsol$ have a median dust temperature of $T_{d}$\,=\,32\,$\pm$\,2\,K. 

The selection function for the observations of local sources is such that the sample is effectively complete in characteristic dust temperature. As discussed above, this is not true for the AS2UDS SMGs as the 850\,$\mu$m selection results in a bias towards cooler sources. Thus, a direct comparison between the low- and high-redshift samples over a wide-range in luminosity is not possible (see Figure~\ref{fig:lumden}). Nevertheless, we can consider the AS2UDS SMG with far-infrared luminosities of $L_{\mathrm{FIR}}$\,=\,3--5$\times$10$^{12}$\,$\Lsol$, where, based on their temperature distribution we would expect to detect sources with the same range in $T_{d}$ as those that are seen in the local Universe. At these luminosities the AS2UDS SMGs have a median characteristic dust temperature of $T_{d}$\,=\,32\,$\pm$\,2\,K, and are thus $\gsim$\,7\,K cooler than the most luminous subset of the local sample, or $\gsim$\,8\,K if we extrapolate the best--fit relation to the local sample to match the median far--infrared luminosity of the AS2UDS SMGs. Including the ALESS SMGs in this comparison raises the median characteristic dust temperature of the SMGs to $T_{d}$\,=\,33\,$\pm$\,1\,K, a negligible difference

As we discuss in the following section the difference in the characteristic dust temperatures of sources at low and high redshift may be due to a difference in the morphology of the dust emitting regions, as more extended sources will result in cooler temperatures, or, alternatively, due to the optical depth of the dust clouds. It is important to stress however that although the difference between the characteristic dust temperatures of the AS2UDS SMGs and low-redshift infrared-bright galaxies may appear modest, it is indicative of a significant difference between the properties of the dust emitting regions in these two galaxy populations, highlighting that SMGs should not be viewed as simple high redshift analogues of local U\,/\,LIRGs (see also \citealt{Chapman04b,Pope08,Hainline09,Ivison10eyelash,Rowlands14}). 

\subsection{Luminosity Density, Brightness Temperature and Optical Depth}
\label{subsec:lumdensity}
A subset of 23 AS2UDS SMGs were detected at a sufficiently high SNR in our 0.3$''$--resolution ALMA imaging to allow a measurement of the intrinsic sizes of their 870\,$\mu$m emission regions (see \citealt{Simpson15}). The far--infrared luminosity surface density of the AS2UDS SMGs as a function of the characteristic dust temperature of each SMG is shown in Figure~\ref{fig:lumden}, where we have assumed that half of the far--infrared luminosity is emitted within the half--light radius of the observed 870\,$\mu$m emission. The data show a clear trend of increasing temperature with luminosity surface density, as expected if warmer dust emission traces regions of increasingly dense star formation.  

However, we must consider that the observed peak of the far--infrared dust SED is sensitive to both the optical depth and the temperature of the dust. As such, the dust temperature derived from SED fitting is dependent on the frequency at which the optical depth is assumed to be unity. Instead, with measured sizes at 870\,$\mu$m, we can determine the brightness temperature ($T_{B}$), a fundamental property of the sources in our sample, by solving:

\begin{equation}
B_{\nu_\mathrm{rest}}(T_B) = 0.5S_{\nu_\mathrm{obs}}(1+z)^3 / \Omega_{\nu_\mathrm{obs}}
\label{eqn:bnu}
\end{equation}

\noindent where $\Omega_{\nu}$\,=\,$\pi R_{\nu}^2$\,/\,D$_\mathrm{A}^{2}$ and represents the solid angle subtended by the source, $R_{\nu}$ is the intrinsic size of the emission region deconvolved from the beam, and a factor of 0.5 is included as we are considering the average brightness temperature within the half--light radius of each source. The brightness temperature represents the temperature of a blackbody radiating at a given luminosity and size. Using Equation\,\ref{eqn:bnu} and the full expression for the Planck function we determine a median brightness temperature for the A2SUDS SMGs of 25\,K, with a 1--$\sigma$ range of 20--32\,K (Table~2). 

The brightness temperature is a fundamental property of each source and can be related to the dust temperature and optical depth as follows 

\begin{equation}
T_B = \frac{h \nu / k}{e^{h \nu / k T_D}} (1-e^{-\tau_{\nu}}).
\label{eqn:tb}
\end{equation}

\noindent where $\tau_{\nu}$\,=\,$(\frac{\nu}{\nu_0})^\beta$ and $\beta$\,=\,1.8 (see \S\,\ref{subsubsec:firfitting}). Thus, with resolved emission at multiple frequencies it is possible to determine the brightness temperature at different wavelengths and hence solve for both the true dust temperature and the optical depth (e.g.\ \citealt{Sakamoto08,Wilson14,Barcos15}). We do not have resolved observations of the AS2UDS SMGs at multiple wavelengths, but we can use the measured size at an observed wavelength of 870\,$\mu$m and the far-infrared\,/\,sub--millimeter photometry from the {\it{Herschel}} and ALMA observations to place constraints on the optical depth. To do so, we assume that half-light radius of the observed emission at all far--infrared wavelengths is the same as that measured at 870\,$\mu$m (see also \citealt{Aravena08, Spilker16}). The emission from each SMG is modeled using Eqn.\,\ref{eqn:tb} and the fitting procedure described in \S\,\ref{subsubsec:firfitting}.  We stress that for a given source the observed size of the emission region is dependent on the optical depth at the emission frequency; optical depth increases with frequency so shorter wavelength emission will appear extended on larger physical scales. As such, our assumption of a constant source size over-estimates the fraction of observed 250, 350 and 500\,$\mu$m flux density that is located within the observed 870--$\mu$m half--light radius, and the optical depth and the true dust temperature derived in our analysis should be considered lower limits.

To determine the optical depth of the SMGs we require that they are: resolved in our 870\,$\mu$m observations; detected in the SPIRE bands; and have a photometric redshift. 14 AS2UDS SMGs satisfy these criteria and we estimate that they have a median optical depth of unity at $\lambda_0$\,(=\,$c$\,/\,$\nu_0$)\,$\ge$\,75\,$\mu$m (1--$\sigma$ dispersion 55--90\,$\mu$m), \footnote{To investigate whether the median optical depth of the AS2UDS SMGs is sensitive to our assumption of $\beta$\,=\,1.8 we repeat our analysis at a fixed value of $\beta$\,=\,1.5 and 2.0 (e.g.\ \citealt{Magnelli12,Dunne01}). We find that the estimated optical depth of each SMG increases systematically with increasing $\beta$, and determine that the median wavelength at which the optical is unity is $\ge$\,$55$\,$\mu$m and $\ge$\,$85$\,$\mu$m for $\beta$\,=\,1.5 and 2.0, respectively. Thus, we caution that our result is mildly sensitive to the assumed value of the dust emissivity and that we adopt a fixed value of $\beta$\=\,1.8 throughout this work, consistent with previous studies.} and a true dust temperature of $T_{d}$\,$>$\,33$^{+3}_{-2}$\,K, a systematic increase of $\ge$\,$3.1^{+1.0}_{-0.3}$\,K relative to the median characteristic dust temperature that we that we measured for the same sources using an optically--thin SED model ($T_{d}$\,=\,30\,$\pm$\,1\,K, where the uncertainty represents the bootstrap error on the median of the sample).

With physically--motivated constraints on the dust temperature and the optical depth we can make an accurate comparison between the luminosity surface density and temperature relation that we determine for the AS2UDS SMGs and the Stefan--Boltzmann law. The Stefan--Boltzmann law between luminosity surface density and temperature is shown in Figure~\ref{fig:lumden}, and represents the expected relation for the faintest source in our sample ($S_{870}$\,=\,3.6\,mJy) with a physical half--light radius of 1.4\,kpc (the median size of the sample; see~\citealt{Simpson15}). While the AS2UDS SMGs follow the trend predicted by the Stefan--Boltzmann law they lie consistently below the predicted relation; an offset that increases linearly as we consider sources brighter than the faintest source in our sample. However, the Stefan--Boltzmann law is only valid for blackbody emission and can only be applied to SMGs if the dust emitting regions are assumed to be optically--thick at all wavelengths. Although, we have shown that the AS2UDS SMGs are optically--thick to at least 75\,$\mu$m we demonstrate that an increase in the optical depth raises the dust temperature of the sources an effect that only increases the offset between the Stefan--Boltzmann law and the SMGs presented here.

We now compare our optical depth measurements for the AS2UDS SMGs to similar studies of far--infrared--bright sources at high and low redshift. Recently, \citet{Riechers13} used extensive FIR-photometry to place an upper limit of $\lambda_0$\,$<$\,163\,$\mu$m and a best--fit value of $\lambda_{0}$\,=\,100$^{+40}_{-60}$\,$\mu$m, on the optical--depth of the high redshift SMG HFLS\,3. While the result presented by \citet{Riechers13} appears to be in agreement with the present work we caution that \citet{Riechers13} consider the total emission from the source, whereas we have determined the typical optical depth within the half--light radius of the sub-millimeter emission. As seen for Arp220, a local ULIRG, the method presented here typically estimates a higher optical--depth for the dust emission when compared to fitting the global far--infrared photometry (see \citealt{Rangwala11,Scoville17}), consistent with the density of dust decreasing radially in these sources. 

The 14 AS2UDS SMGs with optical depth constraints have a median $L_{\mathrm{FIR}}$\,=\,$3.8^{+1.1}_{-0.4}$\,$\times$10$^{12}$\,$\Lsol$ and so class as ULIRGs, despite our earlier analysis showing that they have distinct properties to these local sources. To investigate this discrepancy further we now compare our the optical depth of the AS2UDS SMGs to similarly luminous sources in the local Universe. Interferometric observations of CO or mm--emission have also been used to study selected $z$\,$\approx$\,0 ULIRGs (e.g. \citealt{Scoville91,Downes98,Sakamoto99,Bryant99,Sakamoto08}). These studies have resulted in a consistent picture; that the emission from ULIRGs originates from a compact, $R$\,$\lsim$\,1\,kpc region that has a high column density of molecular gas and is optically--thick at far-infrared wavelengths. For example, \citet{Scoville17} presented high resolution ALMA observations of Arp\,220 at 2.6\,mm. When combined with studies at shorter wavelengths the ALMA observations demonstrate that one of the two nuclei in Arp\,220 is optically--thick to 2.6\,mm, a considerably higher optical depth than the lower limit that we have estimated for the AS2UDS SMGs.

\citet{Lutz16} recently presented a study into the optical depth of 260 far-infrared luminous galaxies at $z$\,$\lsim$\,0.1 with {\it{Herschel}}\,/\,PACS observations at 70 and 160\,$\mu$m. Restricting to sources with a comparable far--infrared luminosity to those studied here ($L_{\mathrm{FIR}}$\,$>$\,10$^{12}$\,$\Lsol$), \citet{Lutz16} show that far-infrared emission from these galaxies arises from regions that are, on average, optically--thin. This comparison suggests that the dust clouds in SMGs are optically--thick at longer wavelengths than local ULIRGs. However, \citet{Lutz16} caution that the estimate of the optical--depth presented in their work should be considered a lower limit if the sources in their sample transition from optically--thick to --thin between 70 and 160\,$\mu$m. We have determined that the AS2UDS SMGs are optically--thick to $\lambda_0$\,=\,75$^{+15}_{-20}$\,$\mu$m, but similarly caution that this wavelength should be considered a lower limit. As such, it is clear that additional high resolution observations of the SMGs presented here, at multiple frequencies, are required to further investigate the difference in the optical depth properties of local ULIRGs and high redshift SMGs.

\subsection{Dust Correction}\label{subsec:dust}
In \S\,\ref{subsec:lumdensity} we showed that the dust emission region in SMGs becomes optically--thick, on average, at a wavelength of $\lambda_0$\,$\ge$\,75$^{+15}_{-20}$\,$\mu$m. The shape, and importantly, the peak wavelength of the far--infrared dust SED are sensitive to both the optical depth and temperature of the dust emission. We have modeled the far--infrared emission from each  SMG with a physically--motivated model that includes an optical depth parameter and find that the dust temperatures are $\ge$\,3.1$^{+1.0}_{-0.3}$\,K higher, on average, relative to the temperature derived from an optically--thin SED. A change in the dust temperature has a direct effect on the derived dust mass of each SMG, with an increase in the dust temperature resulting in a lower total dust mass. We derive dust masses for the 14 AS2UDS SMGs with optical depth constraints and find that they have a median dust mass of $M_{d}$\,=\,6.3$_{-0.8}^{+1.1}$$\times$10$^8$\,$\Msol$. In comparison, the median dust mass of these SMGs in the unphysical optically--thin regime is $M_{d}$\,=\,8.5$_{-0.5}^{+1.8}$$\times$10$^8$\,$\Msol$, which represents an increase of 35\,$\pc$ relative to the optically--thick model. We note that when calculating dust mass we assume a dust opacity of $\kappa_{850{\mu}\mathrm{m}}$\,=\,0.07\,m$^2$kg$^{-1}$ (\citealt{James02}), but that there is likely a factor of three systematic uncertainty in this value (e.g.\ \citealt{James02,Alton04}). When combined with the half-light size of the dust emission the dust masses estimated from out optically--thick SED modeling imply that these SMGs have a median dust surface density of 8.7$_{-0.7}^{+1.3}$$\times$10$^7$\,$\Msol$\,kpc$^{-2}$, within the half--light, or half--mass, radius of $\sim$\,1--2\,kpc 

The high dust column densities that we have determined for the AS2UDS SMGs suggest that optical--to--near--infrared emission from the ongoing starburst will be strongly attenuated, comparable to that observed for local ULIRGs. As shown by \citet{Guver09}, the magnitude of optical attenuation can be related to the column density of hydrogen atoms as follows; 

 \begin{equation}
N_H(\mathrm{cm}^{-2}) = 2.21 \times 10^{21} A_\mathrm{v} (\mathrm{mag}) 
 \end{equation}

\noindent where $A_{\mathrm{v}}$ represents extinction in the rest--frame $V$--band. To estimate the hydrogen column density of the AS2UDS SMGs we first convert the dust mass of each source to a gas mass by adopting a constant gas--to--dust ratio. We follow \citet{Swinbank13}, who present a comparison of the dust masses \citep{Magnelli12} and CO--derived gas masses \citep{Bothwell13} of the same SMGs and suggest that a gas-to-dust ratio of $\delta_{gdr}$\,=\,90\,$\pm$25 is appropriate for SMGs; consistent with the expected gas-to-dust ratio given the metallicity, stellar mass and star formation rates of these sources (\citealt{Draine07,RemyRuyer14}). 

Adopting a gas--to-dust ratio of $\delta_{gdr}$\,=\,90\,$\pm$\,25 we find that the 14 SMGs in our resolved sample have a median hydrogen column density of $N_{H}$\,=\,9.8$_{-0.7}^{+1.4}$$\times$10$^{23}$\,cm$^{-2}$ and thus a median $V$--band dust obscuration of $A_\mathrm{v}$\,=\,540$^{+80}_{-40}$\,mag to the source of the rest--frame $\sim$\,200\,$\mu$m emission. In our analysis we have assumed that the dust is uniformly distributed within the half--light radius, which is consistent with recent high resolution studies of non-lensed SMGs that indicate that the majority of the dust emission originates from a smooth ``disk--like'' component (see \citealt{Hodge16,Oteo16}; Gullberg et al.\ in prep.). We stress that the smooth appearance of the far--infrared emission does not rule out that the dust has an underlying ``clumpy'' morphology (e.g.\ \citealt{Swinbank10Nature,ALMA15}). Instead, we suggest that any ``clumps'' may be embedded in a dust photosphere that is optically--thick, with the far--infrared emission tracing the surface of the optically--thick region. 

The average $V$--band extinction of 540$^{+80}_{-40}$\,mag for the AS2UDS SMGs indicates that effectively all of the optical--to--near--infrared emission from stars that are spatially--coincident with the far--infared emission region will be attenuated by dust. Clearly the magnitude of the attenuation is extreme and we now consider the consequences for the multi-wavelength analysis of SMGs, and hence the physical properties that are determined from optical--to--near--infrared SED fitting (e.g.\ stellar mass and total star formation rate). In SED fitting routines the obscuration by dust is typically modelled by assuming that the dust is distributed in a uniform screen across the galaxy (e.g.\ \citealt{Calzetti00}) or that the obscured star formation occurs in dense clouds embedded within the overall galaxy (e.g.\ {\sc magphys}; \citealt{daCunha08}). A number of studies of interferometrically--identified SMGs have adopted these approaches, with both techniques finding that the detected visible emission from these typical SMGs has a median dust attenuation of $A_{\mathrm{v}}$\,$\sim$\,2\,mag (e.g.\ \citealt{Simpson14,dacunha15}). 

The magnitude of the $V$--band extinction in the AS2UDS SMGs is clearly in strong disagreement with estimates from optical--to--near--infrared SED fitting. To investigate this discrepancy we first consider the applicability of the simple dust screen model to the multi-wavelength analysis of SMGs. Using $H$--band {\it{HST}} imaging, \citet{Chen15} showed that the stellar emission from SMGs has a median half--light radius of 4.4$_{-0.5}^{+1.1}$\,kpc, a factor of $\sim$\,3--4\,$\times$ larger than the dust emission region (see \citealt{Simpson15,Ikarashi15}) and hence an order of magnitude larger in area. Given the discrepancy in the profiles of the dust and the less--obscured stellar emission in SMGs it is unsurprising that a simple dust screen provides a poor representation of the resolved properties of these sources (see \citealt{Hodge16}). Indeed, the discrepancy between the $V$--band attenuation estimated from the dust column density and the simple dust screen can be understood by considering that in SED--fitting the extinction is measured relative to the detectable emission at restframe 1--2\,$\mu$m from a source. As such, the dust--correction is luminosity weighted by the light that is detectable from a source and in the scenario presented here should be considered as a lower limit for each SMG.

We next consider the {\sc magphys} SED-fitting method presented by \citet{daCunha08}. In {\sc magphys} the dust emission from a source is considered to be comprised of two components; a diffuse interstellar medium and compact birth clouds. Energy--balance arguments are then employed to ensure that the emission in the far--infrared, which arises due to the reprocessing of stellar light by dust, is fully consistent with the integrated stellar light of the system. Thus, the energy--balance argument ensures that there is a physically--motivated upper limit to the total stellar emission of the system. However, we have the shown that the far--infrared emission region in SMGs has a typical $V$--band obscuration of $A_\mathrm{v}$\,=\,540$^{+80}_{-40}$\,mag, which corresponds to an optical depth of $\tau_{v}$\,$\sim$\,500$^{+60}_{-30}$. This optical depth is an order of magnitude higher than the model values used in {\sc magphys}, and three orders of magnitude higher than the relatively tight prior that is placed on the parameter values (e.g.\ \citealt{dacunha15}). 

Indeed, the results presented here suggest that any ``energy--balance'' analysis of the emission from SMGs should take into account the spatial distribution of the {\it detectable} optical and far--infrared components; the visible component of the emission and the far--infrared emission by dust should be considered as arising from effectively independent regions of the overall system. However, we strongly caution that in any ``energy--balance'' analysis of SMGs the detectable emission at optical wavelengths will contain no information about the ongoing obscured starburst. Thus, while an ``energy--balance'' analysis will determine a stellar component (including stellar mass) that is consistent with the far--infrared emission we stress that it is determined purely by the prior assumptions on the model parameters. As such, the stellar masses of these systems should be treated with extreme caution (see also \citealt{Hainline11,Michalowski12,Simpson14}).

\subsection{Spheroid Growth?}
The intensity of the ongoing starburst in SMGs, along with their large gas reservoirs that can sustain a prolonged period of stellar mass growth ($\sim$\,100\,Myr; \citealt{Bothwell13}), has led a number of authors to suggest an evolutionary link between these sources and local elliptical galaxies (e.g.\ \citealt{Lilly99, Blain04a, Swinbank06b, Tacconi08, Swinbank10, Simpson14, Toft14}). Furthermore, the recent discovery that the ongoing obscured star formation in SMGs has a half--light radius of $\sim$\,1.5\,kpc has fueled speculation that we are witnessing the direct assembly of a spheroid component in these sources \citep{Simpson15,Ikarashi15,Hodge16}. 

To investigate the possible descendents of the AS2UDS we now estimate the final stellar mass that will be contained within the compact, starburst region. The stellar mass of the ongoing starbursts cannot be estimated through SED--fitting techniques due to the absence of detectable emission at optical--to--near--infrared wavelengths. However, we can constrain the final stellar mass component of the ongoing starburst by considering the gas masses derived for the AS2UDS SMGs with measured sizes at restframe $\sim$\,200\,$\mu$m. Assuming that all of the available gas in these SMGs is converted into stars then the ongoing starburst will result in a compact spheroid with a mass of $\sim$\,1\,$\times$\,10$^{11}$\,$\Msol$. The assumption that all of the gas mass is converted to stellar mass is unrealistic, with large--scale outflows likely removing some gas from the galaxy. However, we have not included any contribution to the final stellar mass from either a pre--existing stellar component or the transformation of the extended stellar component due to a potential ongoing merger in these systems \citep{Chen15}. As such, we consider $\sim$\,10$^{11}$\,$\Msol$ as a reasonable estimate of the stellar mass of the post--starburst remnant.

In the local Universe, the high--mass end of the galaxy stellar mass function ($\gsim$\,1\,$\times$\,10$^{11}$\,$\Msol$) is dominated by elliptical, lenticular (S0), and, to a significantly lesser extent, massive (Sa) spiral galaxies (see \citealt{Kelvin14}). To determine a feasible evolutionary pathway for the AS2UDS SMGs we first investigate whether the estimated spheroid masses are consistent with the properties of local S0 and Sa galaxies, which have typical a spheroid--to--disc mass ratio of $\sim$\,0.4 \citet{Graham08}. We have estimated that the AS2UDS SMGs will form a spheroid component with a stellar mass of $\sim$\,10$^{11}$\,$\Msol$. Thus, if the AS2UDS SMGs are the progenitors of S0 and Sa galaxies then they must correspond to galaxies in the local Universe that have total stellar masses of $\sim$\,2--3\,$\times$\,10$^{11}$\,$\Msol$, where we have neglected to include any stellar mass growth between the SMG--phase and $z$\,=\,0.

S0 and Sa galaxies with total stellar masses of $\sim$\,2--3\,$\times$\,10$^{11}$\,$\Msol$ are extremely rare, with an estimated space density of $\sim$\,10$^{-7}$--10$^{-8}$\,Mpc$^{-3}$ \footnote{We estimate the comoving space density of local S0 \& Sa galaxies by integrating the best--fit Schecter functions to the morphological--type stellar mass functions presented by \citet{Kelvin14}, which were derived from observations taken as part of the Galaxy and Mass Assembly survey (GAMA)} \citep{Kelvin14}. The AS2UDS SMG considered here have a median $870$\,$\mu$m flux density of 8.0\,$\pm$\,0.4\,mJy, corresponding to an estimated comoving space density of $\sim$\,10$^{-5}$\,Mpc$^{-3}$ (\citealt{Karim13,Simpson14,Simpson15b}). Thus, these SMGs are expected to be 2--3 orders of magnitude more numerous than local spirals (Sa) and lenticular galaxies that have spheroid stellar masses that are consistent with these high redshift starbursts. As such, we suggest that {\it if} we are indeed witnessing a centrally--concentrated starburst in SMGs, which is directly growing the stellar mass of a spheroidal component, then they cannot evolve into local spiral or lenticular galaxies (without subsequent significant loss of stellar mass). Instead, SMGs must be the progenitors of todays massive ellipticals and hence are ideal tracers for the formation of the most massive and oldest galaxies at high redshift \citep{Nelan05}.

\section{Conclusions}
In this paper we have presented a comprehensive study of the
multi--wavelength properties of 52 ALMA--identified SMGs in the UDS
field with high spatial resolution 870\,$\mu$m imaging. The main conclusions of our work are:

\begin{itemize}

\item We use the available imaging of the UDS to characterize the properties of the AS2UDS SMGs and show that 35 of the 48 (non--lensed) AS2UDS SMGs (73\,$\pm$\,7\,$\pc$) are detected in the $K$--band imaging. We estimate photometric redshifts for these 35 AS2UDS SMGs and determine that they lie at a median redshift of $z$\,=\,2.65\,$\pm$\,0.13, which rises to $z$\,=\,2.9\,$\pm$\,0.2 if the SMGs with insufficient photometry to derive a photometric redshift are included and are assumed to lie uniformly at $z$\,=\,3--6.

\item We model the far-infrared emission from the sources in our sample and show that they have a median far-infrared luminosity of $L_{\mathrm{FIR}}$\,=\,$3.2^{+0.3}_{-0.7}$\,$\times$10$^{12}$\,$\Lsol$. Combining our AS2UDS sample with the ALESS survey we find that SMGs are $\sim$\,8\,K cooler, at a fixed far-infrared luminosity, compared to local far--infrared--bright galaxies. This is consistent with the larger physical size of the high redshift sources and suggests that SMGs should not be considered scaled-up version of $z$\,$\sim$\,0 ULIRGs.

\item We use a subset of 23 AS2UDS SMGs that have dust emission sizes from high resolution ALMA imaging to constraint the optical depth of the SMG population. The far-infrared sizes show that this emission does not originate from dust clouds that are optically-thin. Instead, we show that the the dust regions in these archetypal SMGs are optically--thick at a median wavelength of $\lambda_0$\,$\ge$\,75\,$\mu$m, with a 1--$\sigma$ dispersion of 55--90\,$\mu$m By modeling the emission with an optically--thick SED we estimate these sources have a true dust temperature that is $\ge$\,$3.1^{+1.0}_{-0.3}$\,K higher than the characteristic dust temperature that is measured by assuming the emission is optically--thin at all wavelengths. Thus, the discrepancy in the characteristic dust temperatures of SMGs and local ULIRGs may be due to SMGs being more optically--thick and larger, or intrinsically cooler and larger.

\item Using the dust masses derived from our physically-motivated, optically-thick SED fits we determine that AS2UDS SMGs have a median hydrogen column density of $N_{H}$\,=\,9.8$_{-0.7}^{+1.4}$\,$\times$\,10$^{23}$\,cm$^{-2}$, corresponding to a median $V$--band obscuration of $A_\mathrm{v}$\,=\,540$^{+80}_{-40}$\,mag averaged along the line of sight to the source of the far--infrared emission. The extreme attenuation in the far--infrared emission region means that effectively all of the stellar light from any co--located stellar component is obscured at optical--to--near--infrared wavelengths. As such, stellar properties that are derived through SED--fitting techniques should be treated with the upmost caution.

\item Finally, we investigate the possible evolutionary pathways for the AS2UDS SMGs. Assuming that the compact, obscured starburst is centrally--concentrated, we estimate that the AS2UDS SMGs will host a post--starburst spheroid with a stellar mass of $\sim$\,10$^{11}$\,$\Msol$. We show that local S0 \& Sa galaxies with a comparable spheroidal mass have a space density that is 2--3 orders of magnitude lower than the AS2UDS SMGs, indicating that SMGs do not evolve into lenticular or spiral galaxies. Instead, we our analysis indicates that SMGs must be the progenitors of local elliptical galaxies.

\end{itemize}

\section*{Acknowledgements}
JMS acknowledges financial support from the ERC Advanced Investigator
projects DUSTYGAL 321334 and COSMICISM 321302. IRS
acknowledges support from the ERC Advanced Investigator program
DUSTYGAL 321334, an RS/Wolfson Merit Award and STFC (ST/L00075X/1). RJI and VA acknowledge financial support from the ERC Advanced Investigator
project COSMICISM 321302. JEG acknowledges support from the Royal Society. We thank Adam Avison and the Manchester ALMA ARC node for their assistance verifying the calibration and imaging of our ALMA data. This work was performed in part at the Aspen Center for Physics, which is supported by National Science Foundation grant PHY-1066293.

This paper makes use of the following ALMA data:
ADS/JAO.ALMA$\#$2012.1.00090.S. ALMA is a partnership of ESO
(representing its member states), NSF (USA) and NINS (Japan), together
with NRC (Canada) and NSC and ASIAA (Taiwan), in cooperation with the
Republic of Chile. The Joint ALMA Observatory is operated by ESO,
AUI/NRAO and NAOJ. This publication also makes use of data taken with
the SCUBA--2 camera on the James Clerk Maxwell Telescope as part of S2CLS. At the time of data acquisition the James
Clerk Maxwell Telescope was operated by the Joint Astronomy Centre on
behalf of the Science and Technology Facilities Council of the United
Kingdom, the National Research Council of Canada, and (until 31 March
2013) the Netherlands Organization for Scientific Research. Additional
funds for the construction of SCUBA--2 were provided by the Canada
Foundation for Innovation. 

All data used in this analysis can be obtained from either the ALMA archive, the Canadian Data Archive Center (CADC\,/\,JCMT), or the WFCAM Science Archive (WSA\,/\,UKIDSS).  

\bibliographystyle{mn2e} 
\bibliography{ref.bib}

% Table of properties
%%
 \begin{table*}
 \centering
 \centerline{\sc Table 1: Observed Properties}
\vspace{0.1cm}
 {%
 \begin{tabular}{lcccccccc}
 \hline
 \noalign{\smallskip}
ID & R.A. & Dec. &  $K^{b}$ & $S_{250}$ & $S_{350}$ & $S_{500}$ & $S^{\rm ALMA}_{870}$ & $S_{1.4\,\mathrm{GHz}}$ \\
   & (J2000) & (J2000) &  (AB) & (mJy) & (mJy) & (mJy) & (mJy) & ($\mu$Jy)  \\  [0.5ex]  
\hline \\ [-1.9ex] 

UDS47.0  &  02:19:24.84 &   $-$05:09:20.7 &   $<$\,24.6 &   $<$\,9.2 &  $<$\,10.6 &   $<$\,12.2 &   8.7\,$\pm$\,0.6 &   85\,$\pm$\,21 \\
UDS47.1  &  02:19:24.64 &   $-$05:09:16.3 &   $<$\,24.6 &   $<$\,9.2 &  $<$\,10.6 &   $<$\,12.2 &   2.1\,$\pm$\,0.8 &   - \\
UDS48.0  &  02:19:24.57 &   $-$04:53:00.2 &   21.49\,$\pm$\,0.02 &  85.2\,$\pm$\,7.8 &  64.5\,$\pm$\,6.7 &  26.3\,$\pm$\,5.1 &  7.5\,$\pm$\,0.5 &   254\,$\pm$\,22 \\
UDS48.1  &  02:19:24.62 &   $-$04:52:56.9 &   22.37\,$\pm$\,0.05 &  $<$\,18.1 &   $<$\,17.0 &   $<$\,17.8 &   1.3\,$\pm$\,0.5 &   67\,$\pm$\,20 \\
UDS57.0  &  02:19:21.14 &   $-$04:56:51.3 &   22.40\,$\pm$\,0.05 &  $<$\,16.7 &   $<$\,16.5 &   $<$\,18.6 &   9.5\,$\pm$\,0.6 &   65\,$\pm$\,21 \\
UDS57.1  &  02:19:20.88 &   $-$04:56:52.9 &   $<$\,24.6 &   27.9\,$\pm$\,4.2 &  36.3\,$\pm$\,5.3 &  37.2\,$\pm$\,6.4 &  5.9\,$\pm$\,0.9 &   - \\
UDS57.2  &  02:19:21.41 &   $-$04:56:49.0 &   25.08\,$\pm$\,0.45 &  $<$\,14.1 &   $<$\,14.9 &   $<$\,18.6 &   1.5\,$\pm$\,0.6 &   - \\
UDS57.3  &  02:19:21.39 &   $-$04:56:38.8 &   $<$\,24.6 &   $<$\,12.3 &   $<$\,13.9 &   $<$\,17.4 &   2.1\,$\pm$\,1.0 &   - \\
UDS74.0  &  02:19:13.19 &   $-$04:47:08.0 &   22.53\,$\pm$\,0.05 &  $<$\,7.7 &  20.1\,$\pm$\,3.9 &  19.4\,$\pm$\,4.1 &  4.5\,$\pm$\,0.5 &   - \\
UDS74.1  &  02:19:13.19 &   $-$04:47:05.0 &   24.24\,$\pm$\,0.23 &  $<$\,9.0 &  $<$\,10.8 &   $<$\,13.9 &   1.5\,$\pm$\,0.5 &   - \\
UDS78.0  &  02:19:09.74 &   $-$05:15:30.6 &   22.82\,$\pm$\,0.08 &  27.3\,$\pm$\,4.1 &  30.7\,$\pm$\,4.9 &  21.6\,$\pm$\,4.4 &  8.2\,$\pm$\,0.5 &   63\,$\pm$\,22 \\
UDS79.0  &  02:19:09.94 &   $-$05:00:08.6 &   22.99\,$\pm$\,0.07 &  $<$\,8.5 &  16.2\,$\pm$\,3.5 &  14.8\,$\pm$\,3.4 &  7.7\,$\pm$\,0.5 &   65\,$\pm$\,17 \\
UDS109.0$^{a}$ &  02:18:50.07 &   $-$05:27:25.5 &   - &   $<$\,9.2 &  $<$\,15.5 &   $<$\,13.9 &   7.6\,$\pm$\,0.7 &   131.5\,$\pm$\,31.8 \\
UDS109.1 &  02:18:50.30 &   $-$05:27:17.2 &   22.23\,$\pm$\,0.04 &  11.4\,$\pm$\,2.4 &  24.2\,$\pm$\,4.4 &  25.5\,$\pm$\,5.0 &  4.2\,$\pm$\,0.6 &   - \\
UDS110.0 &  02:18:48.24 &   $-$05:18:05.2 &   21.17\,$\pm$\,0.02 &  27.0\,$\pm$\,4.1 &  26.4\,$\pm$\,4.6 &  18.6\,$\pm$\,4.0 &  7.7\,$\pm$\,0.6 &   125\,$\pm$\,18 \\
UDS110.1 &  02:18:48.76 &   $-$05:18:02.1 &   21.20\,$\pm$\,0.02 &  20.2\,$\pm$\,3.5 &  20.4\,$\pm$\,4.0 &  $<$\,16.0 &   2.0\,$\pm$\,0.8 &   - \\
UDS156.0 &  02:18:24.14 &   $-$05:22:55.3 &   23.09\,$\pm$\,0.09 &  $<$\,17.8 &   $<$\,17.0 &   $<$\,18.6 &   9.7\,$\pm$\,0.7 &   39.0\,$\pm$\,11.2 \\
UDS156.1 &  02:18:24.24 &   $-$05:22:56.9 &   24.10\,$\pm$\,0.21 &  33.0\,$\pm$\,4.6 &  34.6\,$\pm$\,5.2 &  36.5\,$\pm$\,6.3 &  8.5\,$\pm$\,0.7 &   136\,$\pm$\,45 \\
UDS160.0$^{a}$ &  02:18:23.73 &   $-$05:11:38.5 &   - &  16.5\,$\pm$\,3.1 &  20.6\,$\pm$\,4.0 &  13.0\,$\pm$\,3.1 &  7.9\,$\pm$\,0.6 &   44\,$\pm$\,8 \\
UDS168.0 &  02:18:20.40 &   $-$05:31:43.2 &   21.96\,$\pm$\,0.04 &  $<$\,12.3 &   17.7\,$\pm$\,3.7 &  16.1\,$\pm$\,3.7 &  6.7\,$\pm$\,0.6 &   135\,$\pm$\,32 \\
UDS168.1 &  02:18:20.31 &   $-$05:31:41.7 &   21.96\,$\pm$\,0.04 &  18.3\,$\pm$\,3.2 &  $<$\,16.3 &   $<$\,16.6 &   2.7\,$\pm$\,0.6 &   - \\
UDS168.2 &  02:18:20.17 &   $-$05:31:38.6 &   $<$\,24.6 &   $<$\,11.1 &   $<$\,16.3 &   $<$\,16.6 &   1.5\,$\pm$\,0.7 &   - \\
UDS199.0 &  02:18:07.18 &   $-$04:44:13.8 &   $<$\,24.6 &   $<$\,9.2 &  $<$\,10.8 &   $<$\,12.2 &   4.2\,$\pm$\,0.6 &   68\,$\pm$\,19 \\
UDS199.1 &  02:18:07.19 &   $-$04:44:10.9 &   25.36\,$\pm$\,0.60 &  $<$\,9.2 &  $<$\,10.8 &   $<$\,12.2 &   2.4\,$\pm$\,0.5 &   - \\
UDS202.0 &  02:18:05.65 &   $-$05:10:49.6 &   23.89\,$\pm$\,0.16 &  13.0\,$\pm$\,2.6 &  22.8\,$\pm$\,4.2 &  18.3\,$\pm$\,4.0 &  10.5\,$\pm$\,0.5 &  72\,$\pm$\,16 \\
UDS202.1 &  02:18:05.05 &   $-$05:10:46.3 &   24.27\,$\pm$\,0.22 &  $<$\,8.1 &  $<$\,9.9 &  $<$\,13.9 &   3.5\,$\pm$\,0.9 &   48\,$\pm$\,7 \\
UDS204.0 &  02:18:03.01 &   $-$05:28:41.9 &   23.74\,$\pm$\,0.16 &  $<$\,8.1 &  12.6\,$\pm$\,3.0 &  $<$\,12.9 &   11.6\,$\pm$\,0.6 &  74\,$\pm$\,22\\
UDS204.1 &  02:18:03.01 &   $-$05:28:32.5 &   $<$\,24.6 &   $<$\,9.2 &  $<$\,10.8 &   $<$\,14.5 &   2.2\,$\pm$\,0.9 &   - \\
UDS216.0 &  02:17:56.74 &   $-$04:52:38.9 &   21.01\,$\pm$\,0.01 &  23.6\,$\pm$\,3.8 &  24.6\,$\pm$\,4.4 &  14.5\,$\pm$\,3.4 &  5.2\,$\pm$\,0.5 &   88\,$\pm$\,17 \\
UDS218.0 &  02:17:54.80 &   $-$05:23:23.0 &   22.63\,$\pm$\,0.06 &  16.0\,$\pm$\,2.9 &  15.3\,$\pm$\,3.4 &  21.1\,$\pm$\,4.4 &  6.6\,$\pm$\,0.7 &   58\,$\pm$\,18 \\
UDS269.0$^{a}$ &  02:17:30.44 &   $-$05:19:22.4 &   - &  $<$\,10.0 &   12.8\,$\pm$\,3.1 &  23.2\,$\pm$\,4.7 &  12.9\,$\pm$\,0.6 &  46\,$\pm$\,15 \\
UDS269.1 &  02:17:30.25 &   $-$05:19:18.4 &   22.33\,$\pm$\,0.05 &  12.1\,$\pm$\,2.5 &  $<$\,14.6 &   $<$\,16.8 &   2.0\,$\pm$\,0.7 &   - \\
UDS286.0$^{a}$ &  02:17:25.73 &   $-$05:25:41.2 &   - &  12.1\,$\pm$\,2.5 &  15.0\,$\pm$\,3.4 &  $<$\,18.3 &   5.1\,$\pm$\,0.7 &   103\,$\pm$\,19 \\
UDS286.1 &  02:17:25.63 &   $-$05:25:33.7 &   23.95\,$\pm$\,0.20 &  $<$\,17.6 &   $<$\,16.2 &   $<$\,18.3 &   5.0\,$\pm$\,0.6 &   - \\
UDS286.2 &  02:17:25.80 &   $-$05:25:37.5 &   $<$\,24.6 &   14.1\,$\pm$\,2.8 &  17.5\,$\pm$\,3.7 &  16.0\,$\pm$\,3.7 &  2.6\,$\pm$\,0.6 &   - \\
UDS286.3 &  02:17:25.52 &   $-$05:25:36.7 &   $<$\,24.6 &   $<$\,17.6 &   $<$\,15.7 &   $<$\,18.3 &   1.4\,$\pm$\,0.6 &   - \\
UDS292.0 &  02:17:21.53 &   $-$05:19:07.8 &   22.35\,$\pm$\,0.04 &  17.2\,$\pm$\,3.2 &  13.1\,$\pm$\,3.1 &  17.4\,$\pm$\,3.9 &  4.0\,$\pm$\,0.8 &   52\,$\pm$\,17 \\
UDS292.1 &  02:17:21.96 &   $-$05:19:09.8 &   21.93\,$\pm$\,0.03 &  17.9\,$\pm$\,3.2 &  19.8\,$\pm$\,3.9 &  $<$\,15.7 &   3.6\,$\pm$\,0.8 &   - \\
UDS298.0 &  02:17:19.57 &   $-$05:09:41.2 &   21.83\,$\pm$\,0.03 &  13.9\,$\pm$\,2.7 &  12.7\,$\pm$\,3.0 &  $<$\,13.9 &   1.3\,$\pm$\,0.4 &   - \\
UDS298.1 &  02:17:19.46 &   $-$05:09:33.2 &   22.05\,$\pm$\,0.03 &  $<$\,10.0 &   $<$\,12.6 &   $<$\,12.2 &   1.6\,$\pm$\,0.8 &   - \\
UDS306.0 &  02:17:17.07 &   $-$05:33:26.6 &   21.22\,$\pm$\,1.73 &  53.2\,$\pm$\,6.2 &  36.7\,$\pm$\,5.4 &  $<$\,16.3 &   8.3\,$\pm$\,0.5 &   95\,$\pm$\,22 \\
UDS306.1 &  02:17:17.16 &   $-$05:33:32.5 &   21.31\,$\pm$\,1.89 &  42.4\,$\pm$\,5.5 &  30.4\,$\pm$\,4.9 &  29.1\,$\pm$\,5.4 &  2.4\,$\pm$\,0.4 &   224\,$\pm$\,30 \\
UDS306.2 &  02:17:16.81 &   $-$05:33:31.8 &   $<$\,24.6 &   $<$\,18.1 &   $<$\,16.5 &   $<$\,17.2 &   2.3\,$\pm$\,0.9 &   - \\
UDS334.0 &  02:17:02.47 &   $-$04:57:20.0 &   21.49\,$\pm$\,0.02 &  34.6\,$\pm$\,4.8 &  26.7\,$\pm$\,4.6 &  15.9\,$\pm$\,3.6 &  3.6\,$\pm$\,0.8 &   783\,$\pm$\,16 \\
UDS345.0 &  02:16:57.61 &   $-$05:20:38.6 &   21.47\,$\pm$\,0.02 &  18.0\,$\pm$\,3.2 &  24.5\,$\pm$\,4.4 &  $<$\,15.5 &   2.0\,$\pm$\,0.7 &   74\,$\pm$\,21 \\
UDS361.0 &  02:16:47.92 &   $-$05:01:29.8 &   22.02\,$\pm$\,0.03 &  14.1\,$\pm$\,2.8 &  27.8\,$\pm$\,4.7 &  23.1\,$\pm$\,4.6 &  11.8\,$\pm$\,0.6 &  68\,$\pm$\,22 \\
UDS361.1 &  02:16:47.73 &   $-$05:01:25.8 &   23.64\,$\pm$\,0.15 &  $<$\,9.0 &  $<$\,13.9 &   $<$\,14.8 &   2.0\,$\pm$\,0.7 &   - \\
UDS377.0 &  02:16:41.11 &   $-$05:03:51.4 &   $<$\,24.6 &   14.7\,$\pm$\,2.9 &  16.2\,$\pm$\,3.5 &  $<$\,15.7 &   8.1\,$\pm$\,0.5 &   - \\
UDS392.0 &  02:16:33.29 &   $-$05:11:59.0 &   23.71\,$\pm$\,0.14 &  $<$\,9.2 &  $<$\,11.2 &   $<$\,12.2 &   3.7\,$\pm$\,0.5 &   - \\
UDS408.0 &  02:16:22.26 &   $-$05:11:07.8 &   22.15\,$\pm$\,0.04 &  20.8\,$\pm$\,3.6 &  $<$\,15.9 &   $<$\,13.9 &   9.1\,$\pm$\,0.7 &   101\,$\pm$\,20 \\
UDS408.1 &  02:16:22.28 &   $-$05:11:11.9 &   $<$\,24.6 &   $<$\,10.6 &   20.6\,$\pm$\,4.0 &  15.8\,$\pm$\,3.6 &  2.1\,$\pm$\,0.9 &   - \\
UDS412.0 &  02:16:20.13 &   $-$05:17:26.2 &   $<$\,24.6 &   15.4\,$\pm$\,2.9 &  26.3\,$\pm$\,4.5 &  19.5\,$\pm$\,4.1 &  6.6\,$\pm$\,0.7 &   - \\

 \hline\hline \\  [0.5ex]  
 \end{tabular}
 \begin{flushleft}
 \footnotesize{ $^a$ Identified as a potentially lensed SMG, $^{b}$ Total magnitude.}
 \end{flushleft}
}
 \refstepcounter{table}
 \label{table:obs}
 \end{table*}

%
% Table of properties
%%
 \begin{table*}
 \centering
 \centerline{\sc Table 2: Physical Properties}
\vspace{0.1cm}
 {%
 \begin{tabular}{lcccccccc}
 \hline
 \noalign{\smallskip}
ID  &  $z_{phot}$ & $L^{\mathrm{Thin,~b}}_{\mathrm{FIR}}$ & $T^{\mathrm{Thin,~b}}_{d}$ & FWHM$^c$ & $T_B^{d}$ & $L^{\mathrm{Thick,~e}}_{\mathrm{FIR}}$ & $T^{\mathrm{Thick,~e}}_{d}$ & $\lambda$$_{0}^{e}$ \\
    &             & ($\times$\,10$^{12}$\,$\Lsol$) & (K) &  ($''$) & (K) & ($\times$\,10$^{12}$\,$\Lsol$) & (K) & $\mu$m \\  [0.5ex]  
\hline \\ [-1.9ex] 

UDS47.0  &  - &   - &   - &   0.28\,$\pm$\,0.03 &   - &   - &   - &   - \\
UDS47.1  &  - &   - &   - &   - &   - &   - &   - &   - \\
UDS48.0  &  2.14$^{+0.07}_{-0.15}$ &  11.81$^{+1.25}_{-2.26}$ &   39.7$^{+1.3}_{-2.4}$ &  0.28\,$\pm$\,0.02 &   24.4 &  11.14$^{+1.40}_{-1.82}$ &   45.7$^{+1.7}_{-2.3}$ &  74$^{+7}_{-7}$ \\
UDS48.1  &  2.25$^{+0.15}_{-0.15}$ &  0.05$-$3.08 &   $<$\,44 &   - &   - &   - &   - &   - \\
UDS57.0  &  1.87$^{+0.24}_{-0.17}$ &  0.23$-$0.93 &   $<$\,18 &   0.34\,$\pm$\,0.02 &   20.3 &  - &   - &   - \\
UDS57.1  &  - &   - &   - &   0.26\,$\pm$\,0.05 &   - &   - &   - &   - \\
UDS57.2  &  2.65$^{+0.22}_{-0.31}$ &  0.09$-$3.06 &   $<$\,43 &   - &   - &   - &   - &   - \\
UDS57.3  &  - &   - &   - &   - &   - &   - &   - &   - \\
UDS74.0  &  3.26$^{+0.05}_{-0.11}$ &  3.61$^{+0.43}_{-0.82}$ &  33.5$^{+1.3}_{-2.6}$ &  0.38\,$\pm$\,0.04 &   24.6 &  3.60$^{+0.51}_{-0.70}$ &  36.2$^{+1.7}_{-2.2}$ &  56$^{+9}_{-7}$ \\
UDS74.1  &  4.32$^{+0.37}_{-0.83}$ &  0.27$-$5.85 &   $<$\,55 &   - &   - &   - &   - &   - \\
UDS78.0  &  2.80$^{+0.22}_{-0.19}$ &  5.98$^{+0.90}_{-1.11}$ &  33.1$^{+1.5}_{-2.0}$ &  0.35\,$\pm$\,0.03 &   27.9 &  5.86$^{+1.07}_{-0.91}$ &  37.4$^{+2.1}_{-1.9}$ &  77$^{+8}_{-7}$ \\
UDS79.0  &  3.27$^{+0.07}_{-0.31}$ &  3.43$^{+0.21}_{-1.03}$ &  28.6$^{+0.6}_{-3.2}$ &  0.43\,$\pm$\,0.02 &   27.4 &  3.36$^{+0.39}_{-0.83}$ &  31.6$^{+1.1}_{-2.7}$ &  78$^{+7}_{-11}$ \\
UDS109.0$^{a}$ &  - &   - &   - &   - &   - &   - &   - &   - \\
UDS109.1 &  2.65$^{+0.19}_{-0.09}$ &  2.88$^{+0.50}_{-0.40}$ &  31.3$^{+1.8}_{-1.5}$ &  - &   - &   - &   - &   - \\
UDS110.0 &  1.68$^{+0.24}_{-0.10}$ &  1.60$^{+0.35}_{-0.31}$ &  23.5$^{+1.4}_{-1.4}$ &  0.28\,$\pm$\,0.02 &   19.7 &  1.59$^{+0.42}_{-0.26}$ &  27.5$^{+2.0}_{-1.3}$ &  125$^{+16}_{-13}$ \\
UDS110.1 &  2.80$^{+0.04}_{-0.07}$ &  5.26$^{+1.39}_{-1.12}$ &  44.7$^{+5.5}_{-4.6}$ &  - &   - &   - &   - &   - \\
UDS156.0 &  3.67$^{+0.12}_{-0.13}$ &  1.23$-$4.74 &   $<$\,29 &   0.25\,$\pm$\,0.02 &   55.5 &  - &   - &   - \\
UDS156.1 &  2.35$^{+0.57}_{-0.26}$ &  4.83$^{+1.50}_{-1.47}$ &  30.3$^{+2.6}_{-3.0}$ &  0.24\,$\pm$\,0.03 &   32.4 &  4.84$^{+2.04}_{-1.17}$ &  37.9$^{+4.5}_{-3.1}$ &  120$^{+20}_{-16}$ \\
UDS160.0$^{a}$ &  - &   - &   - &   - &   - &   - &   - &   - \\
UDS168.0 &  2.77$^{+0.06}_{-0.17}$ &  2.65$^{+0.33}_{-0.71}$ &  27.5$^{+1.1}_{-2.7}$ &  0.42\,$\pm$\,0.03 &   22.3 &  2.63$^{+0.41}_{-0.60}$ &  30.1$^{+1.5}_{-2.2}$ &  74$^{+9}_{-10}$ \\
UDS168.1 &  2.77$^{+0.06}_{-0.17}$ &  3.95$^{+0.70}_{-1.02}$ &  40.3$^{+3.3}_{-4.1}$ &  - &   - &   - &   - &   - \\
UDS168.2 &  - &   - &   - &   - &   - &   - &   - &   - \\
UDS199.0 &  - &   - &   - &   0.28\,$\pm$\,0.06 &   - &   - &   - &   - \\
UDS199.1 &  5.01$^{+0.37}_{-2.01}$ &  0.62$-$8.34 &   $<$\,55 &   - &   - &   - &   - &   - \\
UDS202.0 &  3.62$^{+0.44}_{-0.28}$ &  7.06$^{+1.16}_{-1.44}$ &  32.8$^{+1.7}_{-2.4}$ &  0.36\,$\pm$\,0.02 &   39.9 &  7.03$^{+1.51}_{-1.19}$ &  38.6$^{+2.6}_{-2.2}$ &  89$^{+8}_{-9}$ \\
UDS202.1 &  3.35$^{+0.66}_{-0.35}$ &  0.36$-$2.55 &   $<$\,33 &   - &   - &   - &   - &   - \\
UDS204.0 &  3.44$^{+0.59}_{-0.21}$ &  3.33$^{+0.78}_{-0.93}$ &  24.5$^{+1.8}_{-3.0}$ &  0.58\,$\pm$\,0.02 &   26.9 &  3.32$^{+1.12}_{-0.60}$ &  27.2$^{+2.8}_{-1.7}$ &  89$^{+12}_{-12}$ \\
UDS204.1 &  - &   - &   - &   - &   - &   - &   - &   - \\
UDS216.0 &  2.19$^{+0.05}_{-0.09}$ &  2.84$^{+0.33}_{-0.50}$ &  30.3$^{+1.3}_{-1.9}$ &  0.70\,$\pm$\,0.04 &   12.6 &  2.80$^{+0.41}_{-0.42}$ &  31.0$^{+1.7}_{-1.5}$ &  32$^{+4}_{-3}$ \\
UDS218.0 &  3.00$^{+0.17}_{-0.25}$ &  4.02$^{+0.48}_{-0.89}$ &  31.5$^{+1.3}_{-2.5}$ &  0.37\,$\pm$\,0.04 &   26.3 &  3.94$^{+0.62}_{-0.75}$ &  34.7$^{+2.0}_{-2.1}$ &  70$^{+11}_{-9}$ \\
UDS269.0$^{a}$ &  - &   - &   - &   - &   - &   - &   - &   - \\
UDS269.1 &  2.61$^{+0.26}_{-0.10}$ &  2.37$^{+0.76}_{-0.61}$ &  37.9$^{+6.0}_{-4.2}$ &  - &   - &   - &   - &   - \\
UDS286.0$^{a}$ &  - &   - &   - &   - &   - &   - &   - &   - \\
UDS286.1 &  4.91$^{+0.20}_{-0.76}$ &  1.26$-$10.55 &  $<$\,47 &   0.26\,$\pm$\,0.07 &   54.8 &  - &   - &   - \\
UDS286.2 &  - &   - &   - &   - &   - &   - &   - &   - \\
UDS286.3 &  - &   - &   - &   - &   - &   - &   - &   - \\
UDS292.0 &  2.65$^{+0.25}_{-0.07}$ &  3.09$^{+0.59}_{-0.57}$ &  33.5$^{+2.6}_{-2.7}$ &  - &   - &   - &   - &   - \\
UDS292.1 &  2.51$^{+0.23}_{-0.10}$ &  3.03$^{+0.80}_{-0.48}$ &  34.2$^{+3.5}_{-2.4}$ &  - &   - &   - &   - &   - \\
UDS298.0 &  1.81$^{+0.20}_{-0.10}$ &  1.24$^{+0.52}_{-0.32}$ &  34.1$^{+4.9}_{-3.5}$ &  - &   - &   - &   - &   - \\
UDS298.1 &  2.01$^{+0.21}_{-0.18}$ &  0.05$-$1.04 &   $<$\,31 &   - &   - &   - &   - &   - \\
UDS306.0 &  2.31$^{+0.06}_{-0.21}$ &  6.39$^{+0.53}_{-1.39}$ &  33.6$^{+0.9}_{-2.3}$ &  0.30\,$\pm$\,0.02 &   26.1 &  6.15$^{+0.70}_{-1.13}$ &  38.7$^{+1.3}_{-2.2}$ &  86$^{+7}_{-7}$ \\
UDS306.1 &  1.28$^{+0.53}_{-0.06}$ &  2.15$^{+0.84}_{-0.49}$ &  32.9$^{+3.8}_{-2.5}$ &  - &   - &   - &   - &   - \\
UDS306.2 &  - &   - &   - &   - &   - &   - &   - &   - \\
UDS334.0 &  1.93$^{+0.08}_{-0.17}$ &  3.43$^{+0.53}_{-0.94}$ &  34.7$^{+2.3}_{-3.6}$ &  - &   - &   - &   - &   - \\
UDS345.0 &  1.69$^{+0.26}_{-0.05}$ &  1.36$^{+0.47}_{-0.24}$ &  30.0$^{+3.8}_{-2.4}$ &  - &   - &   - &   - &   - \\
UDS361.0 &  3.08$^{+0.18}_{-0.29}$ &  5.51$^{+0.60}_{-1.16}$ &  29.0$^{+1.0}_{-2.1}$ &  0.62\,$\pm$\,0.02 &   23.2 &  5.47$^{+0.77}_{-0.98}$ &  31.3$^{+1.4}_{-1.8}$ &  63$^{+5}_{-5}$ \\
UDS361.1 &  0.61$^{+0.04}_{-0.11}$ &  0.00$-$0.05 &   $<$\,15 &   - &   - &   - &   - &   - \\
UDS377.0 &  - &   - &   - &   0.16\,$\pm$\,0.02 &   - &   - &   - &   - \\
UDS392.0 &  1.72$^{+1.57}_{-0.06}$ &  0.07$-$0.58 &   $<$\,21 &   $<$\,0.18 &   $>$\,22 &   - &   - &   - \\
UDS408.0 &  2.62$^{+0.05}_{-0.13}$ &  3.30$^{+0.34}_{-0.78}$ &  27.4$^{+1.0}_{-2.4}$ &  0.66\,$\pm$\,0.04 &   17.6 &  3.29$^{+0.43}_{-0.66}$ &  28.8$^{+1.3}_{-1.9}$ &  52$^{+6}_{-7}$ \\
UDS408.1 &  - &   - &   - &   - &   - &   - &   - &   - \\
UDS412.0 &  - &   - &   - &   0.30\,$\pm$\,0.07 &   - &   - &   - &   - \\

 \hline\hline \\  [0.5ex]  
 \end{tabular}
 \begin{flushleft}
 \footnotesize{ $^a$ Identified as a potentially lensed SMG, $^{b}$ Assuming an optically--thin SED. The full range of plausible values are given for sources that are only detected in the far-infrared at 870\,$\mu$m. $^{c}$ Intrinsic source size, corrected for synthesized beam, at observed 870\,$\mu$m (see \citealt{Simpson15}) $^d$ Average brightness temperature of the dust contained within the half--light radius of the observed 870\,$\mu$m emission $^e$ Assuming an optically--thick SED, and using observed size of the 870\,$\mu$m emission as a Gaussian prior in the FIR SED fitting.}
 \end{flushleft}
}
 \refstepcounter{table}
 \label{table:phy}
 \end{table*}

\end{document}